\documentclass[prodmode,acmteac]{acmsmall}

\usepackage{amsmath,amssymb,epsfig}
\usepackage{graphicx}

\newcommand{\thm}{\begin{theorem}}
\newcommand{\lem}{\begin{lemma}}
\newcommand{\pro}{\begin{proposition}}
\newcommand{\dfn}{\begin{definition} \rm}
\newcommand{\rem}{\begin{remark}}
\newcommand{\xam}{\begin{example}}
\newcommand{\cor}{\begin{corollary}}
\newcommand{\prf}{\begin{proof}}
\newcommand{\ethm}{\end{theorem}}
\newcommand{\elem}{\end{lemma}}
\newcommand{\epro}{\end{proposition}}
\newcommand{\edfn}{\end{definition}}
\newcommand{\erem}{\bbox\end{remark}}
\renewcommand{\exam}{\bbox\end{example}}
\newcommand{\ecor}{\end{corollary}}
\newcommand{\eprf}{\end{proof}}
\newcommand{\beqn}{\begin{equation}}
\newcommand{\eeqn}{\end{equation}}

\newcommand{\bbox}{\vrule height7pt width4pt depth1pt}
\newcommand{\commentout}[1]{}
\newcommand{\M}{{\cal M}}
\newcommand{\cS}{{\cal S}}

\newcommand{\BR}{\mathit{BR}}

\newenvironment{RETHM}[2]{\trivlist \item[\hskip 10pt\hskip\labelsep{\sc #1\hskip 5pt\relax\ref{#2}.}]\it}{\endtrivlist}
\newcommand{\rethm}[1]{\begin{RETHM}{Theorem}{#1}}
\newcommand{\repro}[1]{\begin{RETHM}{Proposition}{#1}}
\newcommand{\relem}[1]{\begin{RETHM}{Lemma}{#1}}
\newcommand{\recor}[1]{\begin{RETHM}{Corollary}{#1}}

\newcommand{\erethm}{\end{RETHM}}
\newcommand{\erepro}{\end{RETHM}}
\newcommand{\erelem}{\end{RETHM}}
\newcommand{\erecor}{\end{RETHM}}

\DeclareMathOperator*{\argmax}{argmax}
\DeclareMathOperator*{\argmin}{argmin}

\title{An Equilibrium Analysis of Scrip Systems}
\author{IAN A. KASH \affil{Microsoft Research}
ERIC J. FRIEDMAN \affil{University of California, Berkeley}
JOSEPH Y. HALPERN \affil{Cornell University}
}
\begin{abstract}
A game-theoretic model of scrip (artificial currency) systems is analyzed.
It is shown that relative entropy can be used to characterize the
distribution of agent wealth
when all agents use \emph{threshold strategies}---that is, they
volunteer to do work
iff they have below a threshold amount of money.
Monotonicity of
agents' best-reply functions is used
to show that
scrip systems have pure strategy equilibria
where all agents use threshold strategies.
An algorithm is given that can compute such an equilibrium
and the resulting distribution of wealth.
\end{abstract}
\acmformat{}
\begin{document}
\begin{bottomstuff}
Preliminary versions of the material in this paper appeared
    in the Proceedings of the 7th and 8th ACM Conferences on Electronic
    Commerce~\cite{scrip06,scrip07}
    and the Proceedings of the First Conference on Auctions,
    Market Mechanisms and Their Applications~\cite{scrip09}.
Authors' addresses: I. A. Kash, Microsoft Research, 21 Station Road, Cambridge,  UK CB1 2FB; 
E. J. Friedman, International Computer Science Institute and Department of Computer Science, University of California, Berkeley;
and J. Y. Halpern, Computer Science Department, Cornell University, Gates Hall, Ithaca, NY 14853-7501.
\end{bottomstuff}
\maketitle



\section{Introduction}\label{sec:intro}

Historically, non-governmental organizations have issued their own
currencies for a wide variety of purposes.  These currencies, known as
\emph{scrip}, have been used in company towns where government issued
currency was scarce \cite{mining},  in Washington DC to reduce the
robbery rate of bus drivers \cite{busdrivers}, and in Ithaca (New
York) to promote fairer pay and improve the local economy
\cite{hours}.
Scrip systems have also been proposed for a variety of online systems.

To give some examples,
market-based solutions
using scrip systems
have been suggested
for applications such as system-resource allocation \cite{agora},
managing replication and query
optimization in a distributed database \cite{mariposa}, and allocating
experimental time on a wireless sensor network test bed \cite{mirage};
a number of sophisticated scrip
systems have been proposed \cite{gupta03,fileteller02,karma03} to allow
agents to pool resources while
avoiding what is known as \emph{free riding}, where agents
take advantage of the resources the system provides while providing none
of their own (as Adar and Huberman \citeyear{adar00} have shown, this
behavior certainly takes place in systems such as Gnutella);
and
Yootles \cite{yootles} uses a scrip
system as a way of helping groups make decisions
using economic mechanisms
without involving real
money.

In this paper, we provide a formal model in which to analyze scrip
systems.  We describe a simple scrip system and show that, under
reasonable assumptions, for each fixed amount of money there is a
nontrivial equilibrium involving \emph{threshold strategies},%
\footnote{We do not restrict agents to playing threshold strategies, but
our results show it is near optimal for them to do so
when other agents do so as well.}
where an agent accepts a request if he has less than $\$k$
for some threshold $k$. %
Although we refer to our unit of scrip as the dollar,
these are not real dollars, nor do we view ``scrip dollars'' as
convertible to real dollars. 
This is a crucial part of our model, as we assume that scrip is only
valued for its use in the system.

An interesting aspect of our analysis is that, in
equilibrium, the distribution of users with each amount of money is
the distribution that minimizes relative entropy to an appropriate
distribution (subject to the money supply
constraint).  This allows us to use techniques from statistical
mechanics to explicitly compute the distribution of money and thus
agents' best-reply functions.
Using this analysis and results of Tarski~\citeyear{tarski}, 
Topkis~\citeyear{topkis}, and Milgrom and Roberts~\citeyear{MiR90},
we can show that there are pure-strategy equilibria using threshold
strategies, and that these can be found using a simple algorithm.


In a companion paper \cite{scripjournal2}, we use our analysis of
this model to answer questions of interest to system designers.  For
example, we examine how the quantity of money effects the efficiency
of the equilibrium and show that it is maximized by
maintaining the appropriate ratio between the total amount of money
and the number of agents.
This ratio can be found
by increasing the money supply up to the point that the system
experiences a ``monetary crash,'' where money is sufficiently
devalued that no agent is willing to perform a service.
We also incorporate agents altruistically providing service, hoarding
money, creating multiple identities, and colluding into the model.

The rest of the paper is organized as follows.
In Section~\ref{sec:related}, we review related work.  Then
in Section~\ref{sec:model}, we present the formal model.
We analyze the distribution of money in this model when agents are
using threshold strategies in
Section~\ref{sec:distribution}, and show that it is characterized by
relative entropy.
Using this analysis, we show in
Section~\ref{sec:strategic} that, under minimal assumptions, there
is a nontrivial equilibrium where all agents use
threshold strategies.
These results apply to a sufficiently large population of agents after
a sufficiently long period of time, so in
Section~\ref{sec:simulations} we use simulations to demonstrate that
these values are reasonable in practice.
We conclude in Section~\ref{sec:conclusion}.

\section{Related Work}\label{sec:related}

Scrip systems have a long history in computer science, with two
main thrusts: resource allocation and free-riding prevention.
Early applications for resource allocation include allocating time on shared
computers (e.g. the PDP-1 at Harvard University in the
1960s~\cite{sutherland68}), 
agoric systems~\cite{agora}, which envisioned solving problems such as
processor scheduling using markets, and Mariposa~\cite{mariposa}, a
market-driven query optimizer for distributed databases.
More recently, scrip systems have been used to allocate the resources
of research testbeds.  Examples include Mirage~\cite{mirage} for
wireless sensor networks, Bellagio~\cite{auyoung07} for PlanetLab,
and Egg~\cite{egg} for grid computing.
Virtual markets have been used to coordinate the activity of nodes of
a sensor network~\cite{mainland04}.
Yootles~\cite{yootles} uses a scrip to help people make everyday
decisions, such as where to have lunch, without involving real money.

Systems that use scrip to prevent free riding include
KARMA~\cite{karma03}, which provides a general framework for P2P
networks.
Gupta et al.~\citeyear{gupta03} propose what they call a ``debit-credit
reputation computation'' for P2P networks, which is essentially a
scrip system.
Fileteller~\cite{fileteller02} uses payments in a network file storage
system.
Dandelion~\cite{sirivianos07} uses scrip in a content distribution
setting.
Belenkiy et al.~\citeyear{belenkiy07} consider how a BitTorrent-like
system can make use of e-cash.
Antfarm~\cite{antfarm} uses scrip to optimize content distribution
across a number of BitTorrent-like swarms.

Despite this tremendous interest in scrip systems, there has been
relatively little work studying how they behave.
While there has been extensive work in macroeconomics on the effect of
variables such as the amount of money in circulation on the economy
(see, for example, the relevant chapters
of~\cite{blanchard1989lectures}),
this work focuses on goals such as minimizing inflation and maximizing
employment that are not directly relevant to a system designer.

There is also a literature in economics that attempts to understand the
properties of government-backed currencies that are not tied to a
commodity such as gold ({\em fiat money}).  An early model by
Kiyotaki and Wright~\citeyear{kiyotaki89} has some similarity with our
approach in terms of analysis technique, but focuses on money as
useful because it can be held for free, while there are costs associated
with storing a physical good until it can be traded for a different,
desired good.  A large literature has built on this model.  Green and
Zhou~\citeyear{green98} relax Kiyotaki and Wright's assumption that
agents can only 
hold a single unit of money.  In subsequent work, Green and
Zhou~\citeyear{green02} eliminate storage costs in favor of perishable
goods and introduce an analysis of the dynamics of the system.
Perhaps most closely related,
Berentsen~\citeyear{berentsen02a} characterizes the equilibrium
distribution of money in a model with perishable goods.
Although the model is somewhat different, this characterization
corresponds to a simple special case of our characterization.
Berentsen and Rocheteau~\citeyear{berentsen02b} examine the
extent to which inefficiency in this type of model depends on
whether money is indivisible.

An important technical difference in these models from our work is that
there is assumed to be a continuum of agents.  This means that the
evolution of the system is deterministic, and the distribution of money
does not change in steady state.  In contrast, our approach assumes
a large but finite population of agents, and explicitly models the extent
to which fluctuations in the distribution of money occur.
Furthermore, the analysis in this line of work relies on a symmetry
in agent preferences, while our model allows different types of agents
with distinct preferences.  Two
other solution concepts have recently been investigated that look at
the behavior of systems with an infinite number of agents:
\emph{mean field equilibrium}~\cite{adlakha10a,adlakha10b}
and \emph{oblivious equilibrium}~\cite{weintraub08,weintraub10,weintraub11}. 
These solution concepts are similar in spirit to ours, in that they find
equilibria in  
restricted sets of strategies, and then show that these converge to
approximations of the true equilibria.  Our results are somewhat stronger;
we show that the ``restricted equilibrium'' is an approximate
equilibrium, rather than just converging to one.  In fact, we conjecture
that our   
threshold equilibrium is precisely the mean field equilibrium for our
model.

Ng et al.~\citeyear{ng05} studied user behavior in a deployed scrip
system, and observed that users behaved rationally in the presence
of a non-strategyproof combinatorial auction.  Among other manipulations,
they observed users breaking of a single bid into multiple bids to exploit the
greedy nature of the clearing algorithm.  Their observations suggest
that system designers will have to deal with game-theoretic concerns.

Hens et al.~\citeyear{hens} do a theoretical analysis of what can be
viewed as a scrip system in a related model.  There are
a number of significant differences between the models.  First, in the Hens
et al.~model, there is essentially only one type of agent, but an
agent's utility for getting service (our $\gamma_t$) may change over
time.  Thus, at any time, there will be agents who differ in their
utility.
In our model, agents want service only occasionally, so, at each
round, we assume that one agent is chosen (by nature)
to request a service, while other agents decide whether or not
to provide it.  (As the number of agents increases, the time between
rounds decreases, so as to keep this assumption reasonable.)
In the Hens et al.~model, every agent always desires
service, so, at each round, each agent
decides whether to provide service, request service, or opt out, as a
function of his utilities and the amount of money he has.
They assume
that there is no (direct) cost for providing service and everyone is
able to do so.
However, they do assume that agents cannot
simultaneously provide and receive service, an assumption that is
typically unreasonable in a peer-to-peer system.
Rather than having desire for service arise from explicitly modeled
fluctuations in utility, we model it as being exogenously generated
for one agent in each round.
The decrease in time between rounds captures the ability of
agents to provide service at essentially the same time they receive it.
Under this assumption,
a system with optimal performance is one where half the agents request
service and the other half are willing to provide it.
Despite these differences, Hens et al.~also show that agents will use a
threshold strategy.

Aperjis and Johari~\citeyear{aperjis06} examine a model of a P2P
filesharing system as an exchange economy.  They associate a price (in
bandwidth) with each file and find a market equilibrium in the
resulting economy.
Later work by Aperjis et al. does use a currency, but the focus
remains on establishing market prices~\citeyear{aperjis08}
Subsequent to our work,
Dandekar et al.~\citeyear{dandekar11}
examined networks where each agent can issue his own scrip,
Rahman et al.~\citeyear{rahman10} proposed a protocol to automatically
adjust credit policies to maintain stability,
Humbert et al.~\citeyear{humbert11} modeled scrip systems where multiple
agents combine to provide service,
van der Schaar et al.~\citeyear{xu13} studied how to set parameters in
a way that is 
robust to noise, and
Johnson et al.~\citeyear{johnson14} showed that welfare can be
improved if the system prefers volunteers with low amounts of scrip.

The ultimate goal of a scrip system is to promote cooperation.
While there is limited theoretical work on scrip systems, there is a
large body of work on cooperation.
Much of the work involves a large group of agents being randomly
matched to play a game such as prisoner's dilemma.
Such models were studied in
the economics literature~\cite{Kan92,Ell94} and first applied to
online reputations in~\cite{FrR01};
Feldman et al.~\citeyear{Feldman04} apply them to P2P systems.

These models fail to capture important asymmetries in the interactions
of the agents.  When a request is made, there are
typically many people in the network who can potentially satisfy it
(especially in a large P2P network), but not all can. For example,
some people may not have the time or resources to satisfy the
request.  The random-matching process ignores the fact that some
people may not be able to satisfy the request.
(Presumably, if the
person matched with the requester could not satisfy the match, he
would have to defect.) Moreover, it does not capture the fact that
the decision as to whether to ``volunteer'' to satisfy the request
should be made before the matching process, not after.  That is, the
matching process does not capture the fact that if someone is
unwilling to satisfy the request, there may well be others who
can satisfy it. Finally, the actions and payoffs in the prisoner's
dilemma game do not obviously correspond to actual choices that can
be made. For example, it is not clear what defection on the part of
the requester means. Our model addresses all these issues.

Scrip systems are not the only approach to preventing free riding.
Two other approaches often used in
P2P networks are barter and reputation systems.
The essence of barter for our purposes is that when deciding
whether to provide service to others agents only
consider the value created by {\em their own} current or past
interactions with the agent in question.
Perhaps the best-known example of a system that uses barter is
BitTorrent~\cite{cohen},
where clients downloading a file try to find other clients with
parts they are missing so that they can trade, thus creating a
roughly equal amount of work. Since the barter is restricted to
users currently interested in a single file, this works well for
popular files, but tends to have problems maintaining availability
of less popular ones.  An example of a barter-like system built on
top of a more traditional file-sharing system is the credit system
used by eMule~\cite{eMule}. Each user tracks his history of
interactions with other users and gives priority to those he has
downloaded from in the past. However, in a large system, the
probability that a pair of randomly-chosen users will have
interacted before is quite small, so this interaction history will
not be terribly helpful. Anagnostakis and
Greenwald~\citeyear{anagnostakis04} present a more sophisticated
version of this approach, but it still seems to suffer from similar
problems.
More recently, Piatek et al.~\citeyear{piatek08} have proposed a model
based on including intermediaries a single hop away; this model is more
liquid than barter but not as liquid as a full scrip system.

A number of attempts have been made at providing general reputation
systems (e.g. \cite{guha04,gupta03,eigentrust,xiong}).
The basic idea is to
aggregate each user's experience into a global number for each
individual that intuitively represents the system's view of that
individual's reputation. However, these attempts tend to suffer from
practical problems because they implicitly view users as either
``good'' or ``bad'', assume that the ``good'' users will act
according to the specified protocol, and that there are relatively
few ``bad'' users.   Unfortunately, if there are easy ways to game
the system, once this information becomes widely available, rational
users are likely to make use of it. We cannot count on only a few
users being ``bad'' (in the sense of not following the prescribed
protocol). For example, Kazaa uses a measure of the ratio of the
number of uploads to the number of downloads to identify good and
bad users.  However, to avoid penalizing new users, they gave new
users an average rating.  Users discovered that they could use this
relatively good rating to free ride for a while and, once it started
to get bad, they could delete their stored information and
effectively come back as a ``new'' user, thus
circumventing the system (see~\cite{anagnostakis04} for a discussion
and~\cite{FrR01} for a formal analysis of this ``whitewashing'').
Thus, Kazaa's reputation system is ineffective.

\section{The Model}\label{sec:model}

Before specifying our model formally, we give an intuitive description
of what our model captures.
We model a scrip system where, as in a
P2P filesharing system, agents provide each other with service.  There is a
single service (such as file uploading) that agents occasionally want.  In
practice, at any given time, a number of agents will want service but, to
simplify the formal description and analysis, we model the scrip system
as proceeding in a series of rounds where, in each round, a single agent
wants service (the time between rounds will be adjusted to capture the
growth in parallelism as the number of agents grows).%
\footnote{For large numbers of agents, our model converges to one in
which agents
make requests in real time, and the time between an agent's requests
is exponentially distributed.  In addition, the time between requests
served by a single player is also exponentially distributed.}
In each round,
after an agent requests service, other agents have to decide whether
they want to volunteer to provide service.  However, not all agents
may be able to satisfy the request (not everyone has every file).
While, in practice, the ability of agents to provide service at
various times may be correlated for a number of reasons (if I don't
have the file today I probably still don't have it tomorrow; if one
agent does not have a file, it may be because it is rare, so that
should increase the probability that other agents do not have it), for
simplicity, we assume that the events of an agent being able to
provide service in different rounds or two agents being able to provide
service in the same or different rounds are independent.
%
If there is at least one volunteer, someone is chosen from among the
volunteers
(at random)
to satisfy the
request.
Our model allows some agents to be more likely to be chosen (perhaps
they have more bandwidth) but
does not capture rules that take the amount of scrip volunteers have
into account (such rules are considered by Johnson et
al.~\citeyear{johnson14}) or
allow an agent to specify which agent is chosen.
Allowing agents this type of control would break the symmetries we use
to characterize the long-run behavior of the system, and would create new
opportunities for strategic behavior.
%
The requester then gains some utility (he got the file) and
the volunteer loses some utility (he had to use his bandwidth to
upload the file), and the requester pays the volunteer a fee that
we fix at one dollar.  As is
standard in the literature, we assume that agents discount future
payoffs. This captures the
intuition that a reward today is worth more than a reward tomorrow, and
allows us to compute the total utility derived by an agent in an
infinite game.
The amount of utility gained by having a service performed and the
amount lost by performing it, as well as many other parameters may
depend on the agent.

More formally, we assume that agents have a \emph{type} $t$ drawn from
some finite set $T$ of types.
We can describe the entire population of agents
using the pair
$(T,\vec{f})$, where $\vec{f}$ is a vector of length $|T|$ and
$f_t$ is the fraction with type $t$.
In this paper, we consider only what we call \emph{standard
agents}.  These are agents who derive no pleasure from performing a
service, and for whom money 
is valued only for its use in obtaining service.
Thus, for a standard agent, there is no direct connection between money
(measured in dollars) and utility (measured in utils).
We can characterize
the type of a standard agent by a tuple
$t = (\alpha_t, \beta_t, \gamma_t, \delta_t, \rho_t, \chi_t)$, where
\begin{itemize}
\item $\alpha_t > 0$ is the undiscounted cost in utils for an agent of type $t$ to
satisfy a request;
\item $0 < \beta_t < 1$ is the probability that an agent of type $t$ can
satisfy a request;
\item $\gamma_t > \alpha_t$ is the utility that an agent
of type $t$
gains for having a request satisfied;
\item $0 < \delta_t < 1$ is the rate at which an agent of type $t$ discounts
utility;
\item $\rho_t > 0$ represents the (relative) request rate (some people
want files more often than others).
For example, if there are
two types of agents with $\rho_{t_1} = 2$ and $\rho_{t_2} = 1$
then agents of the first type will make requests twice as often as
agents of the second type.  Since these request rates are relative, we
can multiply them all by a constant to normalize them.  To simplify
later notation, we assume the $\rho_t$ are normalized so that
$\sum_{t \in T} \rho_t f_t = 1$;
\item $\chi_t > 0$ represents the (relative) likelihood of an agent to be
chosen when he volunteers (some uploaders may be more popular than
others).
In particular, this means the relative probability of two given agents
being chosen is independent of which other agents volunteer; and
\item $\omega_t = \beta_t \chi_t / \rho_t$ is not part of the tuple,
but is an important derived parameter that, as we will see in
Section~\ref{sec:distribution}, helps determine how much money an
agent will have.
\end{itemize}
We occasionally omit the subscript $t$ on some of these parameters when
it is clear from context or irrelevant.

Representing the population of agents in a system as $(T,\vec{f})$ captures
the essential features of a scrip system we want to model: there are a
large number of agents who may have different types.
Note that some tuples $(T,\vec{f})$ may be incompatible with there being
some number $N$ of agents.  For example, if there are two types, and
$\vec{f}$ says that half of the agents are of each type, then there
cannot be 101 agents.
Similar issues arise when we want to talk about the amount of money in a system
We could deal with this problem in a number of ways (for example,
by having each agent determine his type at random according to the
distribution $\vec{f}$).  For convenience, we
simply do not consider population sizes that are incompatible with $\vec{f}$.
This is the approach used in the literature on
\emph{$N$-replica economies}~\cite{mascolell}.

Formally, we consider games specified by a tuple
$(T,\vec{f},h,m,n)$, where $T$ and $\vec{f}$ are as defined above,
$h \in \mathbb{N}$ is the \emph{base} number of agents of each type,
$n \in \mathbb{N}$ is number of \emph{replicas} of these agents and
$m \in \mathbb{R}^+$ is the
average amount of money.  The total number of agents is thus $hn$.  We
ensure that the
fraction
of agents of type $t$ is exactly $f_t$ and that
the average amount of money is exactly $m$ by requiring that $f_th \in
\mathbb{N}$ and $mh \in \mathbb{N}$.  Having created a base population
satisfying these constraints, we can make an arbitrary number of copies
of it.  More precisely,
we assume that
agents $0 \ldots f_{t_1}h - 1$ have
type $t_1$, agents $f_{t_1}h \ldots (f_{t_1} + f_{t_2})h - 1$ have type
$t_2$, and so on through agent $h - 1$.
These base agents determine the types of all other agents.
Each agent $j \in \{h, \ldots, hn-1\}$ has the same type as $j \mod
h$; that is, all the agents of the form $j + kh$ for $k = 1, \ldots,
n-1$ are replicas of agent $j$.

We also need to specify how money is initially allocated to agents.
Our results are based on the long-run behavior of the system and so
they turn out to hold for any initial allocation of money.  For
simplicity, at the start of the game we
allocate each of the $hmn$ dollars in the system
to an agent chosen uniformly at random,
but all our results would hold if we chose any other initial
distribution of money.

To make precise our earlier informal description,
we describe $(T,\vec{f},h,m,n)$ as an infinite extensive-form game.
A non-root node in the game
tree is associated with a round number (how many requests have
been made so far), a phase number, either 1, 2, 3 , or 4
(which describes how far along we are in determining the results of
the current request),
a vector $\vec{x}$ where $x_i$ is
the current amount of money agent $i$ has, and $\sum_i x_i = mhn$, and,
for some nodes, some
additional information whose role will be made clear below.
We use $\tau(i)$ to
denote the type of agent $i$.

\begin{itemize}

\item The game starts at a special root node, denoted $\Lambda$, where
nature moves.  Intuitively, at $\Lambda$, nature allocates money
uniformly at random, so it transitions to a node of the form
$(0,1,\vec{x})$: round
zero, phase one, and allocation of money $\vec{x}$, and  each possible
transition is
equally likely.

\item At a node of the form $(r,1,\vec{x})$, nature selects an agent
to make a request in the current round.  Agent $i$ is chosen with
probability $\rho_{\tau(i)} / hn$.  If $i$ is chosen, a transition is
made to
$(r,2,\vec{x},i)$.

\item At a node of the form $(r,2,\vec{x},i)$, nature selects the set
$V$ of agents (not including $i$) able to satisfy the request.  Each
agent $j \neq i$ is
included in $V$ with probability $\beta_{\tau(j)}$.  If $V$ is chosen, a
transition is made to $(r,3,\vec{x},i,V)$.

\item At a node of the form $(r,3,\vec{x},i,V)$, each agent in $V$
chooses whether to volunteer.  If $V'$ is the set of agents who choose
to volunteer, a transition is made to $(r,4,\vec{x},i,V')$.

\item At a node of the form $(r,4,\vec{x},i,V')$,
if $V' \neq \emptyset$,
nature chooses a single agent in $V'$ to satisfy the request.
Each agent $j$ is chosen with probability
$\chi_{\tau(j)} / \sum_{j' \in V'} \chi_{\tau(j')}$.  If $j$ is chosen,
a transition is made to $(r+1,1,\vec{x}')$, where
$$x_j' = \left \{
\begin{array}{lll}
x_j - 1 & \mbox{if } i = j \mbox{ and } x_j > 0,\\
x_j + 1 & \mbox{if } j \mbox{ is chosen by nature and } x_i > 0,\\
x_j & \mbox{otherwise.}\\
\end{array} \right.$$
If $V' = \emptyset$, nature has no choice; a transition is made to
$(r+1,1,\vec{x})$ with probability 1.
\end{itemize}

A strategy for agent $\ell$ describes whether or not he will
volunteer at every node of the form $(r,3,\vec{x},i,V)$ such that $\ell \in
V$.  (These are the only nodes where $\ell$ can move.)
We also need to specify what agents know when they make their
decisions.  To make our results as strong as possible, we allow an
agent to base his strategy on the entire history of the game,
which includes, for example, the current wealth of every other agent.
As we show, even with this unrealistic amount of information,
available, it would still be approximately optimal to adopt a simple
strategy that requires little information---specifically, agents need to
know only their current wealth.  That means that our results would
continue to hold as long as agents knew at least this information.
A strategy profile $\vec{S}$ consists of one strategy per agent.
A strategy profile $\vec{S}$ determines a probability
distribution over paths $\Pr_{\vec{S}}$ in the game tree.  Each path
determines the
value of the following random two variables:

\begin{itemize}

\item $x_i^r$, the amount of money agent $i$ has during round $r$,
defined as the value of $x_i$ at the nodes with round number $r$ and

\item $u_i^r$, the utility of agent $i$ for round $r$.
If $i$ is a standard agent, then
$$u_i^r = \left \{
\begin{array}{lll}
\gamma_{\tau(i)} & \mbox{if a node } (r,4,\vec{x},i,V')
\mbox{ is on the path with } V' \neq 0\\
-\alpha_{\tau(i)} & \mbox{if } i \mbox{ is chosen by nature at node }
(r,4,\vec{x},j,V')\\
0 & \mbox{otherwise.}\\
\end{array} \right.$$

\end{itemize}


$U_i(\vec{S})$, the total expected utility of agent $i$ if strategy
 profile $\vec{S}$ is played, is the discounted sum of his
per round utilities $u_i^r$, but the exact form
of the discounting requires some explanation.  In our model, only one
agent makes a request each round.  As the number of agents increases,
an agent has to wait a larger number of rounds to make requests, so
naively discounting utility would mean his utility decreases as the
number of agents increases, even if all of his requests are
satisfied.
This is an artifact of our model breaking time into discrete rounds where
a single agent makes a request.  In reality, many agents make requests
in parallel, and how often an agent desires service typically does not
depend on the number of agents.  It would be counterintuitive to have
a model that says that if agents make requests at a fixed rate and
they are all satisfied, then their expected utility depends on the
number of other agents.  As the following lemma shows, there is a
unique discount rate that removes this dependency.%
\footnote{In preliminary versions of this work we used the discount
  rate of $\delta_t^{1/n}$.  This rate captures the intuitive idea of making
  the time between rounds $1/n$, but results in an agent's utility
  depending on the number of other agents, even if all the agent's
  requests are satisfied.
  However, in the limit as $\delta_t$ goes to 1, agents' normalized
  expected utilities (multiplied by $1 - \delta_t$, as in
  Equation~\ref{eqn:U}) are the same with either discount rate, so our
  main results hold with the discount rate  $\delta_t^{1/n}$ as well.}

\lem
 With a discount rate of $(1 - (1-\delta_t)/n)$, an agent of type
$t$'s expected discounted utility for having all his requests
satisfied is independent of the number of replicas $n$.  Furthermore,
this is the unique such rate such that the discount rate is
$\delta_t$ when $n=1$.
\elem
\prf
The agent makes a request each round with probability $\rho_t / hn$,
so his expected discounted utility for having all his requests
satisfied is 
\begin{align*}
\sum_{r = 0}^{\infty} (1 - (1-\delta_t)/n)^r
(\rho_t \gamma_t / (hn))
&= (\rho_t \gamma_t / (hn)) / (1 - (1 - (1 - \delta_t)/n))\\
&= (\rho_t \gamma_t / h) / (1 - \delta_t)
\end{align*}
This is independent of $n$,
and satisfies
$(1 - (1 - \delta_t)/1) = \delta_t$, as desired.  It is unique because
choosing any other discount rate for some $n$ will cause the value of
the sum to differ from $(\rho_t \gamma_t / h) / (1 - \delta_t)$ for
that $n$.
\eprf

As is standard in economics, for example
in the folk theorem for repeated games~\cite{fudenbergandtiroletext},
we multiply an agent's utility by $(1 - \delta_t)$ so that his
expected utility is independent of his discount rate as well.
With these considerations in mind,
the total expected utility of agent $i$ given the vector of strategies
$\vec{S}$ is
\begin{equation}
\label{eqn:U}
U_i(\vec{S}) = (1 - \delta_{\tau(i)})
\sum_{r = 0}^\infty (1 - (1-\delta_{\tau(i)})/n)^r
E_{\vec{S}}[u_i^r],
\end{equation}

In modeling the game this way, we have implicitly made a number of
assumptions.  For example, we have assumed that all of agent
$i$'s requests that are satisfied give agent $i$ the same utility, and
that prices are fixed.  We discuss the implications of these
assumptions in Section~\ref{sec:conclusion}.

Our solution concept is the standard notion of an approximate Nash
equilibrium.
As usual, given a strategy profile $\vec{S}$ and agent $i$, we use
$(S_i',\vec{S}_{-i})$ to denote the strategy profile that is
identical to $\vec{S}$ except that agent $i$ uses $S_i'$.

\dfn \label{def:bestreply}
A strategy $S_i'$ for agent $i$ is an {\em $\epsilon$-best reply}
to a strategy profile $\vec{S}_{-i}$ for the agents other than $i$ in
the game
$(T,\vec{f},h,m,n)$ if, for all strategies
$S_i''$,
$$U_i(S_i'',\vec{S}_{-i}) \leq
U_i(S_i',\vec{S}_{-i}) + \epsilon.$$
\edfn

\dfn \label{def:equilibrium}
A strategy profile $\vec{S}$  for the game $(T,\vec{f},h,m,n)$
is an \emph{$\epsilon$-Nash equilibrium} if for all agents $i$,
$S_i$ is an $\epsilon$-best reply to $\vec{S}_{-i}$.
A \emph{Nash equilibrium} is an epsilon-Nash equilibrium with $\epsilon=0$.
\edfn

As we show in Section~\ref{sec:strategic},  $(T,\vec{f},h,m,n)$ has
equilibria where agents use a particularly simple type of
strategy, called a \emph{threshold strategy}.
Intuitively, an agent with ``too little'' money will want to work, to
minimize the likelihood of running out due to making a long sequence of
requests before being able to earn more money.
On the other hand, a rational agent with plenty of money
will think it is better to delay working, thanks to discounting.
These intuitions suggest that the agent should volunteer if and only if
he has less than a certain amount of money.
Let $s_k$ be the strategy where an agent
volunteers if and only if the requester has at least 1 dollar and the
agent has less than $k$ dollars.
Note that $s_0$ is
the strategy where the agent never volunteers. While everyone
playing $s_0$ is a Nash equilibrium (nobody can do better by
volunteering if no one else is willing to), it is an uninteresting
one.

We frequently consider the situation where each agent of type $t$
uses the same threshold $s_{k_t}$.  In this case, a single vector
$\vec{k}$ suffices to indicate the threshold of each type, and we can
associate with this vector the strategy
$\vec{S}(\vec{k})$ where
$\vec{S}(\vec{k})_i = s_{k_{\tau(i)}}$ (i.e., agent $i$ of type $\tau(i)$
uses threshold $k_{\tau(i)}$).

For the rest of this paper, we focus on threshold strategies (and show
why it is reasonable to do so).
In particular, we show that, if all other agents use threshold
strategies, it is approximately optimal for an agent to use one as
well.  Furthermore there exist Nash equilibria where agents do so.%
\footnote{These equilibria actually satisfy the stronger condition of
(approximate) subgame perfection.} 
While there are potentially other equilibria that use different
strategies, if a system designer has agents use threshold strategies by
default (e.g., through the standard behavior of the client software),
no agent will have an incentive to change.
Since threshold strategies have such low information requirements, they
are a particularly attractive choice for a system designer
as well for the agents, since they are so easy to play.

For  threshold strategy $\vec{S}(\vec{k})$, if $mhn \geq \sum_t f_t k_t hn$
the system will quickly reach
a state where each agent has $k_t$ dollars, so no agent will
volunteer.  This is equivalent to all agents using $s_0$, and
similarly uninteresting.  Therefore, in our analysis we assume that
$mhn < \sum_t f_t k_t hn$.

\section{Analyzing the Distribution of Wealth}\label{sec:distribution}

Our main goal is to show that there exists an approximate
equilibrium where all agents play threshold strategies.  In this
section, we examine a more basic question: if all agents play a
threshold strategy, what happens?
We show that
there is some distribution over money
(i.e., a distribution that describes what fraction of people have each
amount of money) such that the
system ``converges'' to this distribution in a sense to be made
precise shortly.

To motivate our interest in the distribution, consider an agent $i$ who is
trying to decide on a best response in a setting where all other agents
are playing a threshold strategy, and all agents of a particular type
play the same strategy.  Specifically, suppose that agent $i$
has \$$k$, and is trying to decide whether to volunteer.  Assume that
the system is sufficiently large that one agent's decision does not
affect the distribution.   To figure out whether to volunteer, agent $i$
must compute how likely he is to run out of money before he gets a
chance to make another dollar.  Clearly this depends on how much money
he has.  But it also depends on how likely he
is to be chosen when he volunteers, which, in turn, depends on how many
other volunteers there are
(and thus on how many agents are not at their threshold).
Our results show that there is a
distribution of money $d^*$ such that, with extremely high probability, the
actual distribution is almost always extremely close to $d^*$.  By
knowing $d^*$, the agent will know what fraction of agents of each
type $t$ have each amount of money.  If all the agents of type $t$
use the same threshold strategy, he will also know how
many agents of type $t$ volunteer.  Moreover, this number will be
essentially the same at every round.  This will enable him to figure out
when he should volunteer.

We remark that, in addition to providing an understanding of system
behavior that underpins our later results, this result also provides
a strong guarantee about the stability of the economy.
It shows that we do not have wild swings of behavior; in particular, the
fraction of agents volunteering is essentially constant.%
\footnote{Our results show that such swings occur with extremely small
probability.  While this guarantees they will eventually occur, the
expected time this takes is so large that it will effectively never
happen in practice for even a moderately sized system.  Even if such an
unlikely event did occur, our results guarantee that it is transient and that
the system will converge back toward the steady-state distribution.} 

Suppose that all agents of each type $t$ use the same threshold $k_t$, so
we can write the vector of thresholds as $\vec{k}$.
For simplicity, assume that each agent has
at most $k_t$ dollars.  We can make this assumption with
essentially no loss of generality, since if someone has more than $k_t$
dollars, he will just
spend money until he has at most $k_t$ dollars.  After this point he
will never acquire more than $k_t$. Thus, eventually the system will
be in a state where, for all types $t$, no agent of type $t$ has more
than $k_t$ dollars.

We are interested in the vectors $\vec{x}^r$ that can be observed in
round $r$ (recall that $x_i^r$ is the amount of money that agent $i$ has
at round $r$).
By assumption, if agent $i$ has type $\tau(i)$, then $x_i^r \in
\{ 0, \ldots, k_{\tau(i)} \}$.  In addition, since
the total amount of money is $hmn$,
$$\vec{x}^r \in X_{T,\vec{f},h,m,n,\vec{k}} =
\{ \vec{x} \in \mathbb{N}^{hn} \mid \forall i.x_i \leq k_{\tau{i}},
\sum_{i} x_i = hmn \}.$$

The evolution of $\vec{x}^r$ can be described by a Markov chain
$\M_{T,\vec{f},h,m,n,\vec{k}}$ over
the state space $X_{T,\vec{f},h,m,n,\vec{k}}$.
For brevity, we refer to the Markov chain and state space as $\M$
and $X$, respectively, when the subscripts are clear from context.
It is possible to move from state $s$ to state $s'$
in a single round if, by choosing a particular agent $i$ to make
a request and another agent $j$ to satisfy it, $i$'s amount of money in
$s'$ is 1 more than in $s$; $j$'s amount of money in $s'$ is 1 less than
in $s$', and all other agents have the same amount of money in $s$ and $s'$.
Therefore,
the probability of a transition from a
state $\vec{x}$ to $\vec{y}$ is 0 unless there exist two agents $i$
and $j$ such
that $\vec{y}_{i'} = \vec{x}_{i'}$ for all $i' \notin \{i,j\}$,
$ \vec{y}_i = \vec{x}_i + 1$, and $\vec{y}_j = \vec{x}_j - 1$.
In this case, the probability of transitioning from $\vec{x}$ to
$\vec{y}$ is the probability of $j$ being chosen to make a request and
%
$i$ being chosen to satisfy it.
Let $\Delta_{\vec{f},m,\vec{k}}$ denote the set of probability
distributions $d$ on $\cup_{t \in T} \{t\} \times \prod_t \{ 0,
\ldots, k_t \}$ such that for all
types
$t$, $\sum_{l = 0}^{k_t} d(t,l) = f_t$ 
and $\sum_t \sum_{l = 0}^{k_t} l d(t,l) = m.$
We can think of $d(t,l)$ as the fraction of agents
that are of type $t$ and have $l$ dollars.
We can associate each state $\vec{x}$ with its corresponding
distribution $d^{\vec{x}}$.
This is a useful way of looking at the system, since we typically
just care about the fraction of people with each amount of money, not
the amount that each particular agent has.
We show that, if $n$ is large, then there is a distribution $d^* \in
\Delta_{\vec{f},m,\vec{k}}$ such that, after a sufficient amount of
time, the Markov chain $\M$
is almost always in a state $\vec{x}$ such that $d^{\vec{x}}$
is close to $d^*$. Thus, agents can base their decisions about what
strategy to use on the assumption that they will be in a state where the
distribution of money is essentially $d^*$.
Note that, since agents discount future utility, the transient
behavior of the Markov chain does matter, but
by making $\delta_t$ sufficiently large (i.e., if agents are
sufficiently patient) the effect on utility can be made arbitrarily small.
Similarly, for sufficiently large $n$, the effect on utility due to
extremely rare deviations from $d^*$ becomes arbitrarily small.

We can in fact completely characterize the distribution $d^*$. Given
two distributions $d,q \in \Delta_{\vec{f},m,\vec{k}}$, let
$$H(d||q) = \sum_{(t,j) s.t. q(t,j) \neq 0} d(t,j) \log
\left( \frac{d(t,j)}{q(t,j)} \right) $$
denote the \emph{relative entropy} of $d$ relative to $q$ ($H(d||q) =
\infty$ if $d(t,j) = 0$ and $q(t,j) \neq 0$ or vice versa); this is
also known as the \emph{Kullback-Leibler divergence of $q$ from
$d$}~\cite{cover}.
If $\Delta$ is a closed convex set of distributions, then it is well
known that, for each $q$, there is a unique distribution in $\Delta$
that minimizes the relative entropy to $q$.
Since $\Delta_{\vec{f},m,\vec{k}}$ is easily seen
to be a closed convex set of distributions,
in particular, this is the case for $\Delta_{\vec{f},m,\vec{k}}$.
We now show that there exists a $q$ such that, for $n$
sufficiently large, the Markov chain $\M$
is almost always in a state $\vec{x}$ such that $d^{\vec{x}}$ is close
to the distribution
$d^*_{q,\vec{f},m} \in \Delta_{\vec{f},m,\vec{k}}$ that
minimizes entropy relative to $q$.
(We omit
some or all of
the subscripts on $d^*$ when they are not relevant.)
The
statement is correct under a number of senses of ``close''.  For
definiteness, we consider the Euclidean distance.
Given $\varepsilon > 0$
and $q$, let $X_{T,\vec{f},h,m,n,\vec{k},\varepsilon,q}$
(or $X_{\varepsilon,q}$, for brevity)
denote the set of
states $\vec{x} \in X_{T,\vec{f},h,m,n,\vec{k}}$
such that $\sum_{(t,j)} |d^{\vec{x}}(t,j) - d^*_q|^2 <
\varepsilon$.

Let $I^r_{q,n,\varepsilon}$ be the random variable that is 1 if
$d^{\vec{x}^r} \in X_{\varepsilon,q}$, and 0 otherwise.

\thm \label{thm:distribution}
For all games $(T,\vec{f},h,m,1)$, all vectors $\vec{k}$ of thresholds,
and  all $\varepsilon > 0$,
there exist $q \in \Delta_{\vec{f},m,\vec{k}}$ and $n_\varepsilon$ such that,
for all $n > n_\varepsilon$,
there exists a round $r^*$  such that, for all $r > r^*$,
we have $\Pr(I^r_{q,n,\varepsilon} = 1) > 1-\varepsilon$. \ethm

The proof of Theorem~\ref{thm:distribution} can be found in
Appendix~\ref{sec:entropy}.
One interesting special case of the theorem is when
there exist $\beta$, $\chi$, and $\rho$ such that for all types $t$,
$\beta_t = \beta$, $\chi_t = \chi$, and $\rho_t = \rho$.  In this case
$q$ is the distribution $q(t,j) = f_t / (k_t + 1)$ (i.e., $q$ is
uniform within each type $t$).
We sketch the
proof for this special case here.

\prf (Sketch)
Using standard techniques, we can show that our Markov Chain has
a  \emph{limit distribution} $\pi$ such
that for all $\vec{y}$, $\lim_{r \rightarrow
\infty} \Pr(\vec{x}^r = \vec{y}) = \pi(\vec{y})$.
Let $T_{\vec{x}\vec{y}}$ denote the probability
of transitioning from (recurrent) state $\vec{x}$ to (recurrent) state $\vec{y}$.
It is easily verified by an explicit computation of the transition
probabilities that (in this special case)
$T_{\vec{x}\vec{y}} = T_{\vec{y}\vec{x}}$
It is well known that this symmetry implies that $\pi$ is the uniform
distribution \cite{Resnick}.
Thus, after a sufficient amount of time, the
distribution of $\vec{x}^r$ will be arbitrarily close
to uniform.

Since, for large $r$, $\Pr(\vec{x}^r = \vec{y})$ is approximately
1 / $|X|$, the
probability of $\vec{x}^r$ being in a set of
states is the size of the set divided by the total number of
states.
Using a straightforward combinatorial argument, it
can be shown that the fraction of states not in
$X_{\varepsilon,q}$
is bounded by $p(n) / e^{cn}$, where $p$ is a
polynomial.  This fraction goes  to 0 as $n$ gets large.
Thus,  for sufficiently large $n$,
$\Pr(I^r_{q,n,\varepsilon} = 1) > 1 - \varepsilon$.
\eprf

The last portion of the proof sketch is actually a standard technique
from statistical mechanics that involves showing that there is a
\emph{concentration phenomenon} around the maximum entropy
distribution \cite{Jaynes}.
To illustrate what we mean by a concentration phenomenon,
consider a system with
only two dollars.  With $n$ agents, there are $O(n^2)$ ways to assign
the dollars to different agents and $O(n)$ ways to assign them to the
same agent.
If each way of assigning the two dollars to agents is equally likely, we
are far more likely to see a distribution of money where two agents have
one dollar each than one where a single agent has two dollars.
In the special case considered in the proof sketch, when $\pi$ is the
uniform distribution, the number of states corresponding to a
particular distribution $d$ is proportional to $e^{nH(d)}$ (where $H$
here is the standard entropy function).  In general, each state is not
equally likely, which is why the general proof in
Appendix~\ref{sec:entropy} uses relative entropy.\footnote{
Note that in generalizing to relative entropy we switch from
maximizing to minimizing; maximizing entropy is
equivalent to minimizing relative entropy relative to the uniform
distribution.}

Theorem~\ref{thm:distribution} tells us that, after enough time, the
distribution of money is almost always close to some $d^*$, where
$d^*$ can be characterized as a distribution that minimizes relative
entropy subject to
some constraints.
In Appendix~\ref{sec:entropy}, we show that the appropriate
distribution $q$ is
$q(t,i) = (\omega_t)^i / (\sum_t \sum_{j = 0}^{k_t} (\omega_t)^j)$.
The following lemma shows how we can compute $d^*$ from $q$.

\lem \label{lem:minrelent}
\begin{equation}
\label{eqn:d}
d^*(t,i) = \frac{f_t \lambda^i q(t,i)}
{\sum_{j = 0}^{k_t} \lambda^j q(t,j)},
\end{equation}
where $\lambda$
is the unique value such that
\begin{equation}
\label{eqn:m}
\sum_t \sum_i i d^*(t,i) = m.
\end{equation}
\elem

The proof of Lemma~\ref{lem:minrelent} is omitted because it can be
easily checked using Lagrange multipliers in the manner
of~\cite{Jaynes} where the function to be minimized is the relative
entropy of $d^*$ relative to $q$ and the constraints are that an $f_t$
fraction of the agents are of type $t$ and the average amount of money
is $m$.

To give some intuition for the form of $d^*$, suppose that there is only
a single agent who randomly receives opportunities to earn and spend
money, but receives opportunities to earn $\omega_t$ times as often as
opportunities to spend, is unwilling to earn more than $k_t$ dollars,
and is unable to spend when he has zero dollars.  Up to a normalizing
constant, his steady state probability of having $i$ dollars is
$q(t,i)$.  In the full system, each agent is not walking independently,
but instead is limited by the common amount of money.  This acts as a
common bias on their walks, a factor which is captured by $\lambda$. 

\section{Existence of Equilibria}\label{sec:strategic}

We have seen that the system is well behaved if the agents all
follow a threshold strategy; we now want to show that,
if the discount factor $\delta$ is sufficiently large for all agents,
there is a
nontrivial approximate Nash equilibrium where they do so (that is, an
approximate Nash
equilibrium where all the agents use $s_k$ for some $k > 0$.)
To understand why we need $\delta$ to be sufficiently large, note that
if $\delta$ is small, then agents have
no incentive to work.  Intuitively, if future utility is
sufficiently discounted, then all that matters is the present, and
there is no point in volunteering to work.
Thus, for sufficiently small $\delta$, $s_0$ is the only equilibrium.
%
To show that there is a nontrivial equilibrium if the discount factor is
sufficiently large, we first show that, if every other agent
is playing a threshold strategy, then there is an approximate best
reply that is also a threshold strategy.
Furthermore, we show that the best-reply function is monotone; that is,
if some agents change their strategy to one with a higher
threshold, no other agent can do better by lowering his threshold.
This makes our game one with what Milgrom and Roberts~\citeyear{MiR90}
call \emph{strategic complementarities}.
Using results of Tarski~\citeyear{tarski}, Topkis~\citeyear{topkis}
showed that there are pure strategy equilibria in
such games, since the process of starting with a strategy profile
where everyone always volunteers (i.e., the threshold is $\infty$)
and then iteratively computing the best-reply profile to it
converges to a Nash equilibrium in pure strategies.  This
procedure also provides an efficient algorithm for explicitly
computing equilibria.



To see that threshold strategies are approximately optimal,
consider a
single agent $i$ of type $t$ and fix the vector $\vec{k}$ of thresholds
used by the other agents.  If we assume that the number of agents is
large, what an agent $i$ does has essentially no effect on the
behavior of the system (although it will, of course, affect that agent's
payoffs).  In particular, this means that the distribution $q$ of
Theorem~\ref{thm:distribution} characterizes the distribution of money
in the system
with high probability.
(This applies only after some period of time, but for patient agents the
importance of this initial period is negligible.)
This distribution, together with the vector $\vec{k}$ of
thresholds, determines what fraction of agents volunteers at each step.
This, in turn, means that from the perspective of agent $i$, the problem
of finding an optimal response to the strategies of the other agents
reduces to finding an optimal policy in a Markov decision process
(MDP)
$\mathcal{P}_{G,\vec{S}(\vec{k}),t}$.  The behavior of the MDP
$\mathcal{P}_{G,\vec{S}(\vec{k}),t}$ depends on
two probabilities: $p_u$ and $p_d$.  Informally,
$p_u$ is the probability of $i$ earning a dollar during each round if
is willing to volunteer, and $p_d$ is the probability that $i$ will be chosen
to make a request during each round.  Note that $p_u$ depends on
$m$, $\vec{k}$, and $t$
(although it turns out that $p_d$ depends only on $n$, the
number of agents in the system);
if the dependence of $p_u$ on
$m$, $\vec{k}$, and/or $t$ is important, we add the relevant parameters
to the superscript, writing, for example, $p_u^{m,\vec{k}}$.
We show that the optimal policy for $i$ in
$\mathcal{P}_{G,\vec{S}(\vec{k}),t}$ is a threshold policy, and
that this policy is an $\varepsilon$-optimal strategy for $G$.
Importantly, the same policy is optimal independent of the value of $n$
(as long as $n$ is sufficiently large).
This allows us to ignore the exact size of the system in our
later analysis.

For our results,
it will be important to
understand how $p_u$, $p_d$, and $t$ affect the optimal policy for
$\mathcal{P}_{G,\vec{S}(\vec{k}),t}$, and thus the $\varepsilon$-optimal
strategies in the game.  We use this understanding in this section to
show that there exist nontrivial equilibria in
Lemma~\ref{lem:nontrivial} and for a number of results in our
companion paper~\cite{scripjournal2}.

In the following lemma, whose proof
(and the relevant formal definitions)
are deferred to Appendix~\ref{sec:threshold},
Equation~(\ref{eqn:policy}), quantifies the effects of these
parameters.  When choosing whether he should volunteer with his
current amount of money, an agent faces a choice of whether to pay a
utility cost of $\alpha_t$ now in exchange for a discounted payoff of
$\gamma_t$ when he eventually spends the resulting dollar.  His choice
will depend on how much time he expects to pass before
he spends that dollar
(captured by the random variable $J$ in equation~\ref{eqn:policy}),
which in turn depends on his current amount of
money $k$ and the probabilities $p_u$ and $p_d$.
The following lemma
quantifies this calculation.

\lem \label{lem:br}
Consider the games $G_n = (T,\vec{f},h,m,n)$
(where $T$, $\vec{f}$, $h$, and $m$ are fixed, but $n$ may vary).
There exists a $k$ such that for all $n$, $s_k$ is
an optimal policy for $\mathcal{P}_{G_n,\vec{S}(\vec{k}),t}$.
The threshold
$k$ is the maximum value of $\kappa$ such that
\begin{equation}
\label{eqn:policy}
\alpha_t \leq E[(1 - (1 - \delta_t)/n)^{J(\kappa,p_u,p_d)}] \gamma_t,
\end{equation}
where $J(\kappa,p_u,p_d)$ is a random variable whose value is the first
round in which an agent starting with $\kappa$ dollars,
using strategy $s_\kappa$, and with probabilities $p_u$ and $p_d$ of earning
a dollar and of being chosen given that he volunteers,  respectively,
runs out of money.
\elem

Note that the uniqueness of the optimal value of $k$ holds generically.  In
particular, it holds unless \eqref{eqn:policy} is satisfied with equality.

The following theorem shows that an optimal threshold policy for
$\mathcal{P}_{G,\vec{S}(\vec{k}),t}$ is an $\varepsilon$-optimal
strategy for $G$.  In particular, this means that
Equation~(\ref{eqn:policy}) allows us to understand how changing
parameters affect an $\varepsilon$-optimal strategy for $G$, not just
for $\mathcal{P}_{G,\vec{S}(\vec{k}),t}$.

\thm \label{thm:threshold}
For all games $G = (T,\vec{f},h,m,n)$, all vectors $\vec{k}$ of
thresholds, and  all $\varepsilon > 0$,
there exist $n^*_\varepsilon$ and $\delta^*_{\varepsilon,n}$
such that for all $n > n^*_\varepsilon$,
types $t \in T$, and $\delta_t > \delta^*_{\varepsilon,n}$,
an optimal threshold policy for
$\mathcal{P}_{G,\vec{S}(\vec{k}),t}$ is an
$\varepsilon$-best reply
to the strategy profile $\vec{S}(\vec{k})_{-i}$
for every agent $i$ of type $t$.
\ethm

We defer the proof of Theorem~\ref{thm:threshold} to
Appendix~\ref{sec:threshold}.
While, in this and later theorems,
the acceptable
values of $\delta^*_{\varepsilon,n}$
depend on $n$, they are independent if, as we
suggest in Section~\ref{sec:simulations}, the Markov Chain from
Section~\ref{sec:distribution} is rapidly mixing.


Given a game $G = (T,\vec{f},h,m,n)$ and a vector $\vec{k}$ of
thresholds, Lemma~\ref{lem:br} gives an optimal threshold $k_t'$ for
each type $t$.  Theorem~\ref{thm:threshold} guarantees that $s_{k_t'}$
is an $\varepsilon$-best reply to $\vec{S}_{-i}(\vec{k})$, but does
not rule out the possibility of other best replies.  However, for
ease of exposition, we will call $k_t'$ \emph{the} best reply
to $\vec{S}_{-i}$ and call
$\BR_G(\vec{k}) = \vec{k}'$ the best-reply function.  The following
lemma shows that this function is monotone (non-decreasing).
Along the way, we prove that several other quantities are monotone.
First, we
show that $\lambda_{m,\vec{k}}$, the value of $\lambda$ from
Lemma~\ref{lem:minrelent} given $m$ and $\vec{k}$, is non-decreasing
in $m$ and non-increasing in $\vec{k}$.  We use this to show
that $p_u^{m,\vec{k}}$ is non-increasing in $\vec{k}$, which is needed
to show the monotonicity of $\BR_G$.  We defer the proof to
Appendix~\ref{sec:threshold}.

\lem \label{lem:monotone}
Consider the family of games $G_m = (T,\vec{f},h,m,n)$ and the
strategies $\vec{S}(\vec{k})$, for $mhn < \sum_t f_t k_t hn$.
For this family of games,
$\lambda_{m,\vec{k}}$ is non-decreasing in $m$ and non-increasing in
$\vec{k}$; $p_u^{m,\vec{k}}$ is non-decreasing in
$m$ and non-increasing in $\vec{k}$; and the function
$\BR_G$ is non-decreasing in $\vec{k}$ and non-increasing in
$m$.
\elem


The intuition behind monotonicity is easy to explain: if the agents
other than agent $i$ use a higher threshold, then they will volunteer
more often.  Thus, agent $i$ is less likely to be chosen when he
volunteers, and thus he will need to volunteer more often (and so use a
higher threshold himself).
Monotonicity is enough to guarantee the existence of an equilibrium.
We actually know that there is an equilibrium even without the use of
monotonicity.  If all agents choose a threshold of \$0, so no
agent ever volunteers, then clearly $i$'s best response is also never to
volunteer; getting a dollar is useless if it can never be spent.
Fortunately, we can use monotonicity to show that there is a nontrivial
equilibrium in threshold strategies as well.
Indeed, to guarantee the existence of a nontrivial
equilibrium, it suffices to show there is some vector $\vec{k}$ of
thresholds such that $\BR_G(\vec{k}) > \vec{k}$.%
\footnote{We write $\vec{k}' \geq \vec{k}$ to denote $k_t' \geq k_t$ for
all $t$, with strictness if the inequality for at least one $t$ is
strict.}  The 
following lemma, whose proof is again deferred to
Appendix~\ref{sec:threshold}, shows that we can always find such a
point for sufficiently large $\delta_t$.

\lem \label{lem:nontrivial}
For all games $G = (T,\vec{f},h,m,n)$, there exists
a $\delta^*<1$ such
that if $\delta_t > \delta^*$ for all $t$, there is a vector
$\vec{k}$ of thresholds such that $\BR_G(\vec{k}) > \vec{k}$.
\elem

We are now ready to prove our main theorem: there exists a non-trivial
equilibrium where all agents play threshold strategies greater than zero.

\thm \label{thm:equilib}
For all games $G = (T,\vec{f},h,m,1)$ and all $\epsilon$, there exist
$n^*_\epsilon$ and $\delta^*_{\epsilon,n}$
such that, if $n > n^*_\epsilon$ and
$\delta_t > \delta^*_{\epsilon,n}$
for all $t$, then there exists a nontrivial vector $\vec{k}$ of thresholds
that is an $\epsilon$-Nash equilibrium.
Moreover, there exists a greatest such vector
(i.e. an equilibrium where the threshold used by each type is weakly
higher than in every other equilibrium). 
\ethm

\prf
By Lemma \ref{lem:monotone},
$\BR_G$ is a non-decreasing function on a complete lattice, so
Tarski's fixed point
theorem~\cite{tarski} guarantees the existence of a
greatest and least fixed point; these fixed points are equilibria.
The least fixed point is the trivial equilibrium.
We can compute the greatest fixed point by starting with the
strategy profile $(\infty, \ldots, \infty)$ (where each agent uses
the strategy $S_\infty$ of always volunteering) and considering
\emph{$\epsilon$-best-reply dynamics},
that is, iteratively computing the $\epsilon$-best-reply
strategy profile.  Monotonicity guarantees this process converges to
the greatest fixed
point, which is an equilibrium (and is bound to be an equilibrium
in pure strategies, since the best reply is always a pure
strategy).
Since there is a finite amount of money, this process needs to be
repeated only a finite number of times.
By Lemma~\ref{lem:nontrivial}, there
exists a $\vec{k}$ such that $\BR_G(\vec{k}) > \vec{k}$.  Monotonicity
then guarantees that $\BR_G(\BR_G(\vec{k})) \geq \BR_G(\vec{k})$ and
similarly for any number of applications of $\BR_G$.
If $\vec{k}^*$ is the greatest fixed point of $\BR_G$,
then $\vec{k}^* > \vec{k}$.  Thus, the greatest fixed point is a
nontrivial equilibrium.
\eprf

The use of Tarski's fixed point theorem in the proof of Theorem
\ref{thm:equilib} also provides an algorithm for 
finding equilibria that seems efficient in practice: start with the
strategy profile $(\infty, \ldots, \infty)$ and iterate the best-reply
dynamics until an equilibrium is reached.

\begin{figure}[htb]
\centering \epsfig{file=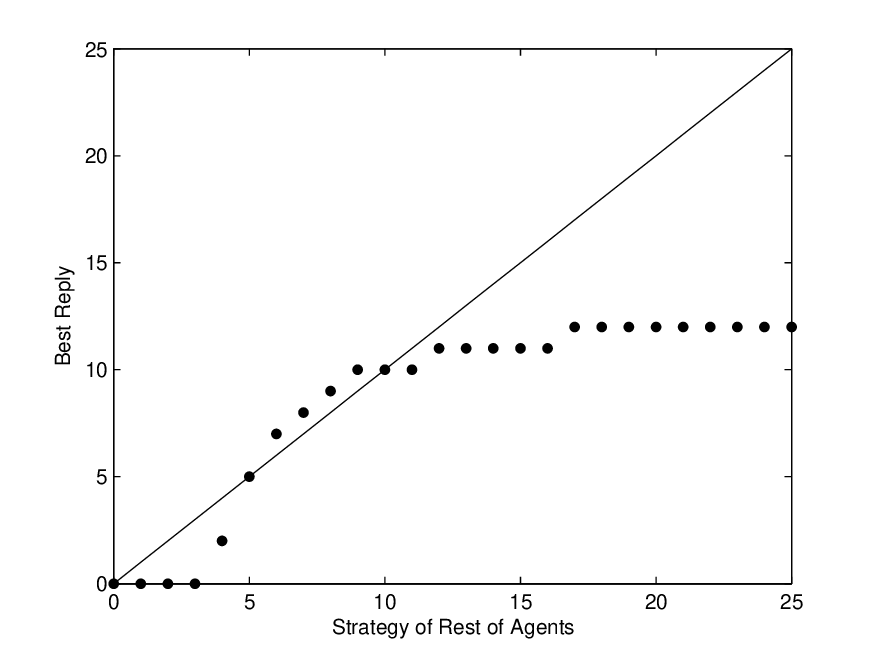, height=2.2in}
\caption{A hypothetical best-reply function with one type of agent.}
\label{fig:br}
\end{figure}

There is a subtlety in our results.
In general, there may be many equilibria.  From the perspective of
social welfare, some will be better than others.
As we show in our companion paper, strategies that use smaller (but
nonzero) thresholds increase social welfare
(intuitively, when agents are less willing to accumulate money, there are
fewer agents with zero dollars and so fewer missed opportunities). 
Consider the best-reply function
shown in Figure~\ref{fig:br}.
In the game $G$ in the example, there is only one type of agent, so
 $\BR_G: \mathbb{N}  \rightarrow \mathbb{N}$.
In equilibrium, we must have $\BR(k) = k$; that is, an
equilibrium is characterized by
a point on the line $y = x$.  This example has three
equilibria, where all agents play  $s_0$, $s_5$, and $s_{10}$
respectively.
The strategy profile where all agents play
$s_5$ is the equilibrium that maximizes social welfare,
while $s_{10}$ is the greatest equilibrium.

In our companion paper, we focus on the greatest equilibrium
in all our applications (although a number of our results hold
for all nontrivial equilibria).  This equilibrium has several
desirable properties.  First, it is guaranteed to be stable;
best-reply dynamics from nearby points converge to it.  By way of contrast,
best-reply dynamics moves the system away from the equilibrium $S_5$ in
Figure~\ref{fig:br}.  Unstable
equilibria are difficult to find in practice, and seem unlikely to be
maintained for any length of time.  Second, the ``greatest'' equilibrium
is the one found by the natural algorithm given in
Theorem~\ref{thm:equilib}.  The proof of the theorem shows that it is also
the outcome that will occur if agents adopt the
reasonable initial strategy of starting with a large threshold and then
using best-reply dynamics.
Finally, by focusing on the
worst nontrivial equilibrium, our results provide
guarantees on social welfare, in the same way that results
on \emph{price of anarchy} \cite{roughgarden02} provide guarantees
(since price of anarchy considers the social welfare of the Nash
equilibrium with the worst social welfare).

\section{Simulations} \label{sec:simulations}

Theorem~\ref{thm:distribution} proves that, for a sufficiently large
number $n$ of agents,
and after a sufficiently large number $r$ of rounds, the distribution of wealth
will almost always be close to the distribution that minimizes
relative entropy.  In this section, we simulate the game to gain an
understanding of how large $n$ and $r$ need to be in practice.
The simulations show that our theoretical results apply even to relatively
small systems; we get tight convergence with a few thousand agents, and
weaker convergence for smaller numbers, in very few rounds
rounds, indeed, a constant number per agent.

\begin{figure}[htbp]
\centering \epsfig{file=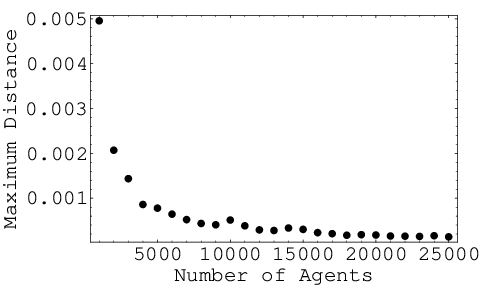, width=3.25in}
\caption{Maximum Euclidean distance from minimum relative entropy
distribution $d^*$ over $10^6$ timesteps.} \label{fig:MaxDist}
\end{figure}

The first simulation explores the tightness of convergence to the
distribution that minimizes relative entropy for various values of
$n$.  We used a single type of agent, with
$\beta = \rho = \chi = 1$, $m = 2$, and $k = 5$.
 For each value of
$n$, the simulation was started with a distribution of money as close
as possible to the distribution $d^*$.  (Recall that $d^*$ is the distribution
that minimizes relative entropy to the distribution $q$ defined in
Theorem~\ref{thm:distribution}, and that $d^*$
characterizes the distribution of money in equilibrium
when the threshold strategy 5 is used.)

We then computed the maximum Euclidean distance between $d^*$ and the
observed distribution over $10^6$ rounds.
As Figure~\ref{fig:MaxDist} shows, the system
does not move far from $d^*$ once
it is there.  For example, if
$n=5000$, the system is never more than distance $.001$ from $d^*$.
If $n=25,000$, it is never more than
$.0002$ from $d^*$.

\begin{figure}[htbp]
\centering \epsfig{file=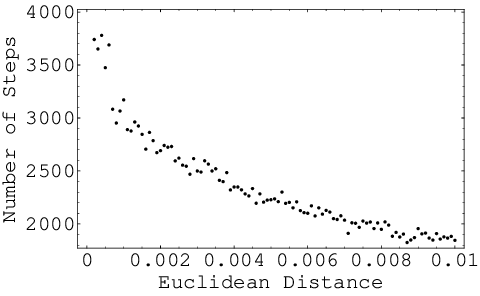, width=3.25in}
\caption{Distance from minimum relative entropy distribution with 1000
agents.} \label{fig:EforN}
\end{figure}

Figure~\ref{fig:MaxDist} does show a larger distance for $n=1000$,
although in absolute terms it is still small.  The next simulation
shows that, while the system may occasionally move away from
$d^*$, it quickly converges
back to it.
We averaged 10 runs of the Markov chain, starting from an
extreme distribution (every agent has either \$0 or \$5), and
considered the average time needed to come within various distances
of $d^*$.  As Figure~\ref{fig:EforN}
shows, after 2 rounds per agent, on average, the Euclidean distance from
the average distribution of money to $d^*$
is .008; after 3 rounds per agent, the distance is down to
.001.

\begin{figure}[htbp]
\centering \epsfig{file=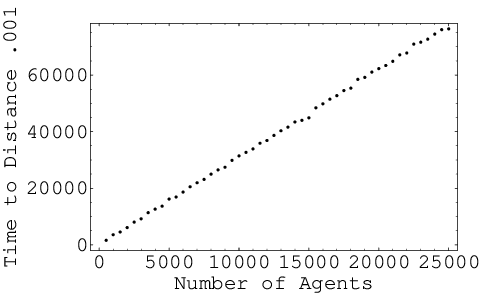, width=3.25in}
\caption{Average time to get within .001 of the minimum relative entropy
distribution.} \label{fig:NforE}
\end{figure}

Finally, we considered more carefully how quickly the system
converges to $d^*$ for various values of
$n$. There are approximately $k^n$ possible states, so the
convergence time could in principle be quite large.  However, we
suspect that the Markov chain that arises here is \emph{rapidly
mixing}, which
means that it will converge significantly faster
(see~\cite{lovasz95} for more details about rapid mixing).  We believe
that the actually time needed is $O(n)$. This behavior is
illustrated in Figure~\ref{fig:NforE}, which shows that for our
example chain (again averaged over 10 runs), after approximately $3n$
steps, the Euclidean distance between the actual distribution of money
in the system  and $d^*$ is less than
.001.  This suggests that we should expect the system to converge in a
constant number of rounds per agent.

Note that this analysis assumes that agents follow the same strategy even
when the system has a distribution of money far from $d^*$.  If agents
are aware that the system is in such a state and can find a better strategy,
then convergence would not necessarily be so rapid.  However, we would
never expect to observe such a deviation in practice in even a moderately
large system, and therefore these results help give confidence that the
deviations we would expect to see (for which the equilibrium strategies
will still be approximately optimal) will be short-lived.  Even if such
a large 
deviation were to somehow occur (e.g., if a bug led to unusual behavior)
this rapid convergence means that the loss due to suboptimal play during
that time would be small.  As there is not an obvious profitable deviation,
it seems reasonable to believe that players would continue to follow
equilibrium behavior.

\section{Discussion} \label{sec:conclusion}

We have given a formal analysis of a scrip system and have shown
that approximate equilibria exist in threshold
strategies and that the distribution of money in these equilibria is
given by relative entropy.  As part of our equilibrium argument, we
have shown that the best-reply function is monotone.  This proves the
existence of equilibria in pure strategies and permits efficient
algorithms to compute these equilibria.

Our model makes a number of assumptions that are worthy of further
discussion.  Some of the simplifying assumptions
can be relaxed without significant changes to our results
(albeit at the cost of greater strategic and analytic
complexity).
At a high level, our results show the system converges to a
steady state when agents follow simple threshold strategies
and that
there is in fact an equilibrium in these strategies.
If, for example, rather than all requests having the same
value to agent ($\gamma_t$), the value of a request is stochastic, agents
might wish to have thresholds for each type of request.
This would allow an agent to forgo a low-valued request if he is low
on money. This makes the space of agent strategies larger and
significantly complicates the proofs in the appendix, but
this high-level characterization still holds.

The most significant assumption we make is that prices are fixed.
However, our results provide insight even if we relax this assumption.
With variable prices, the behavior of the system depends on the value
of $\beta$, the probability that an agent can satisfy a request.  For
large $\beta$, where are a large number of agents who can satisfy
each request, we expect the resulting
competition to effectively produce a fixed price, so our analysis
applies directly.  For small $\beta$, where there are few volunteers
for each request, variable prices can have a significant impact.

However, allowing prices to be set endogenously, by bidding, has a number
of negative consequences.  For one thing, it
removes the ability of the system designer to optimize the system using
monetary policy. In addition, for small $\beta$, it is possible
for colluding agents to form a cartel to fix prices on a resource they
control.  It also greatly increases the strategic complexity of using
the system: rather than choosing a single threshold, agents need an
entire pricing scheme.
Finally, the search costs and costs of executing a transaction
are likely to be higher with floating prices.
Thus, we believe that adopting a fixed price or
a small set of fixed prices is often a reasonable design decision.



\appendix

\section{Proof of Theorem~\ref{thm:distribution}} \label{sec:entropy}

Given a Markov chain $\M$ over a state space $X$ and
state $s \in \cS$, let $I^r_{\vec{x},\vec{y}}$ be the random variable
that is 1 if $\M$ is in state $\vec{y}$ at time $r$ and the
chain started in state $\vec{x}$ and 0 otherwise.
Then $\lim_{r \rightarrow \infty} \Pr(I^r_{\vec{x},\vec{y}}=1)$ is the
limit probability of being in state $\vec{y}$
given that the Markov chain starts in state $\vec{x}$.
In general, this limit does not exist.  However, there are well-known
conditions under which the limit exists, and is independent of
the initial state $\vec{x}$. A Markov chain is said to be
\emph{irreducible} if every state is reachable from every other
state; it is \emph{aperiodic} if, for every state $\vec{x}$, there exist
two cycles from $\vec{x}$ to itself such that the gcd of their lengths is
1.

\thm\label{thm:limit} \text{\cite{Resnick}} If $\M$ is a finite,
irreducible, and aperiodic Markov chain over state space $X$, then
there exists a $d: X \rightarrow \mathbb{R}$ such that, for all $\vec{x}$ and $\vec{y} \in X$,
$\lim_{r \rightarrow \infty} \Pr(I^r_{\vec{x},\vec{y}}=1) = d(\vec{y})$. \ethm

Thus, if we can show that $\M$ is finite, irreducible, and aperiodic,
then the limit distribution exists and is independent of the start
state $\vec{x}$.  This is shown in the following lemma.

\lem
If there are at least three agents, then
$\mathcal{M}$ is finite, irreducible, and aperiodic and
therefore has a limit distribution $\pi$. \elem

\prf
$\M$ is clearly finite since
$X$ is finite.
We prove that it is irreducible by showing that state $\vec{y}$ is
reachable from state $\vec{x}$ by induction on the distance
$w = \sum_{i = 1}^{n} |x_i - y_i|$ (i.e., the sum of the absolute
differences in the amount of money each person has in states $\vec{x}$
and $\vec{y}$).  If $w = 0$,
then $\vec{x} = \vec{y}$ so we are done.
Suppose that $w > 0$ and all pairs of
states that are less that $w$ apart are reachable from each other.
Consider a pair of states $\vec{x}$ and $\vec{y}$ such that the
distance from $\vec{x}$ to $\vec{y}$ is $w$.
Since $w > 0$ and
the total amount of money is the same in all states, there must exist
$i_1$ and $i_2$ such that $x_{i_1} > y_{i_1}$ and $x_{i_2} < y_{i_2}$.
Thus, in state $\vec{y}$, $i_1$ is willing to work (since he has
strictly less than the threshold amount of money) and $i_2$ has money
to pay him (since $i_2$ has a strictly positive amount of money).  The
state $\vec{z}$ that results from $i_1$ doing work for $i_2$ in state
$\vec{y}$ is
of distance $w-2$ from $\vec{x}$. By the induction hypothesis,
%
$\vec{z}$ is reachable from $\vec{x}$.
Since $\vec{y}$ is clearly reachable  from $\vec{z}$,
$\vec{y}$ is reachable from $\vec{x}$.

Finally, we must show that $\M$ is aperiodic.
Suppose $\vec{x}$ is a state such that there exist three agents $i_1$, $i_2$,
and $i_3$ where $i_1$ has more than 0 dollars and $i_2$ and $i_3$ have
less than their threshold amount of money.
There must be such a state by our assumption that
$mhn < \sum_t f_t k_t hn$.
Clearly there is a cycle of length 2 from $\vec{x}$ to itself:
$i_2$ does work for $i_1$ and then $i_2$ does work for $i_1$.  There is
also a cycle of length 3: $i_2$ does work for $i_1$, $i_3$ does work for
$i_2$, then $i_1$ does work for $i_3$.
By irreducibility, identifying a single state with this property is sufficient.
\eprf

We next give an explicit formula for the limit distribution.
Recall that in the special case discussed in the main text, $\beta_t$,
$\chi_t$, and $\rho_t$ were the same for all types, so the transition
probabilities were symmetric and the limit distribution was uniform.
While with more general values they are no longer symmetric, they
still have significant structure that allows us to give a concise
description of the limit distribution.

\lem \label{lem:stationary}
For all states $\vec{x}$ of $\M$,
let $w_{\vec{x}} = \prod_{i} ( \beta_{\tau(i)} \chi_{\tau(i)} /
\rho_{\tau(i)})^{x_i}$, and let $Z = \sum_{\vec{y}} w_{\vec{y}}$.
Then the
limit distribution of $\M$ is $\pi(\vec{x}) = w_{\vec{x}} / Z$. \elem

\prf
Define $\pi$ by taking $\pi(\vec{x}) = w_{\vec{x}} / Z$, where
$w_{\vec{x}}$ and $Z$ are as in the statement of the lemma.
If $T_{\vec{x}\vec{y}}$ is the probability of transitioning from state
$\vec{x}$ to state $\vec{y}$,
it is well known that it suffices to show that $\pi$ satisfies the
\emph{detailed balance condition}~\cite{Resnick}, i.e.,
$\pi(\vec{x}) T_{\vec{x}\vec{y}} = \pi(\vec{y})
T_{\vec{y}\vec{x}}$ for all states $\vec{x}$ and $\vec{y}$
and $\pi$ is a probability measure.
The fact that $\pi$ is a probability measure is immediate from its
definition.
To check
the first condition, let $\vec{x}$ and $\vec{y}$ be adjacent states such that
$\vec{y}$ is reached from $\vec{x}$ by $i$ spending a dollar and $j$ earning a
dollar.  This means that for the transition from $\vec{x}$ to
$\vec{y}$ to happen, $i$ must be chosen
to spend a dollar and $j$ must be able to work and chosen to earn
the dollar.  Similarly for the reverse transition to happen, $j$ must
be chosen to
spend a dollar and $i$ must be able to work and chosen to earn the
dollar.  All other agents have the same amount of money in each
state, and so will make the same decision in each state.  Thus the
probabilities associated with each transition differ only in the
relative likelihoods of $i$ and $j$ being chosen at each point.
These may differ for three reasons: one might be more likely to be
able to satisfy requests ($\beta$), to want to make requests ($\rho$),
or to be chosen to satisfy requests ($\chi$).  Thus,
for some $p$, which captures the effect of other agents volunteering
on the likelihood of $i$ and $j$ being chosen,
we can write the transition probabilities as
$T_{\vec{x}\vec{y}} = \rho_{\tau(i)} \beta_{\tau(j)}
\chi_{\tau(j)} p$ and $T_{\vec{y}\vec{x}} =
\rho_{\tau(j)} \beta_{\tau(i)} \chi_{\tau(i)} p$.
From the definition of $\pi$, we have that
$$\frac{\pi(\vec{x})}{\pi(\vec{y})}
= \frac{\beta_{\tau(i)} \chi_{\tau(i)} \rho_{\tau(j)} }{ \rho_{\tau(i)}
\beta_{\tau(j)} \chi_{\tau(j)}} =
\frac{T_{\vec{y}\vec{x}}}{T_{\vec{x}\vec{y}}}.$$
Thus,
$\pi(\vec{x}) T_{\vec{x}\vec{y}} = \pi(\vec{y}) T_{\vec{y}\vec{x}}$, as
desired.  \eprf

Note that for the special case considered in the main text,
Lemma~\ref{lem:stationary} shows that the limit distribution is the
uniform distribution.

The limit distribution tells us the long run probability of being in a
given state.  Theorem~\ref{thm:distribution} does not mention states
directly, but rather the distributions of money associated with a
state.
In order to prove the theorem, we need to know the probability
of being in some state associated with a given distribution.
This is established in the following lemma.

\lem \label{lem:count}
Let $\pi$ be the limit distribution from Lemma~\ref{lem:stationary},
and let
$V(d) = H(d) - H(\vec{f}) - \log Z +
\sum_t \sum_{i = 0}^{k_t} i d(t,i) \log \omega_t$
(where $H$ is the standard entropy function; that is,
$H(d) = \sum_{t,i} d(t,i) \log d(t,i)$).
For all $d \in \Delta_{\vec{f},m,\vec{k}}$,
either $\pi(\{\vec{x} \mid d^{\vec{x}} = d \}) = 0$ or
$F(hn)e^{hnV(d)} \leq  \pi(\{\vec{x} \mid d^{\vec{x}} = d \})
\leq G(hn)e^{hnV(d)}$,
where $F$ and $G$ are polynomials.
\elem

\prf
Before computing the probability of being in such a state, we first
compute the number of states.
It is possible that there is no state $\vec{x}$ such that
$d = d^{\vec{x}}$ (e.g., if $hn$ is odd and $d$ has half the agents
with 0 dollars).  If there is such a state $\vec{x}$, each such
state has $hnd(t,i)$ agents of type $t$ with $i$ dollars.
Thus, the number of states
$\vec{x}$ with $d = d^{\vec{x}}$
is the number of ways
to divide
the agents into groups of these sizes.  Since there are $hnf_t$ agents
of type $t$, the number of such states is
$$\prod_t {hnf_t \choose hnd(t,0), \ldots, hnd(t,k_t)}.$$
To complete the proof, we use the fact (shown in the proof of Lemma
3.11 of \cite{MaxEnt}) that
$$\frac{1}{F(hn)} e^{hnf_tH(d_t)} \leq {hnf_t \choose hnd(t,0), \ldots,
hnd(t,k_t)} \leq G(hn) e^{hnf_tH(d_t)},$$
where $F$ and $G$ are polynomial in $hn$, and $d_t$ is the distribution
restricted to a single type $t$ (i.e., $d_t(i) = d(t,i) / \sum_i
d(t,i)$).
The \emph{(generalized) grouping property} \cite{cover} of entropy
allows us to express $H(d)$ in terms of the entropy of the
distributions for each fixed $t$, or the $H(d_t)$.  Because
$f_t = \sum_{i} d(t,i)$, this has the particularly simple form
$H(d) = H(\vec{f}) + \sum_t f_t H(d_t)$.
Thus,
up to a polynomial factor,
the number of such states is
$$\prod_t e^{hnf_tH(d_t)}
= e^{hn (\sum_t f_t H(d_t))}
= e^{hn(H(d) - H(\vec{f})}.$$

By Lemma~\ref{lem:stationary}, each of theses states has the same
probability $\pi(\vec{x})$.  Thus, dropping the superscript $\vec{x}$
on $d^{\vec{x}}$ for brevity, the probability of being in such a state
is:
%
\begin{eqnarray*}
 e^{hn(H(d) - H(\vec{f}))}\pi(\vec{x})
& = & e^{hn(H(d)-H(\vec{f}))} \prod_i (\beta_i\chi_t / \rho_i)^{x_i} / Z\\
& = & e^{hn(H(d)-H(\vec{f}))} Z^{-hn} \prod_i (\omega_{t_i})^{x_i}\\
& = & e^{hn(H(d)-H(\vec{f}))} Z^{-hn} \prod_t
\prod_{i = 0}^{k_t} (\omega_t)^{hn i d(t,i)}\\
& = & e^{hn(H(d)-H(\vec{f}))} Z^{-hn} \prod_t \prod_{i = 0}^{k_t}
e^{hn i d(t,i) \log \omega_t}\\
& = & e^{hn(H(d)-H(\vec{f}) - \log Z + \sum_t \sum_{i = 0}^{k_t}
i d(t,i) \log \omega_t)}\\
& = & e^{hnV(d)}
\end{eqnarray*}
\eprf

%
Theorem~\ref{thm:distribution} says that there exists a $q \in
\Delta_{\vec{f},m,\vec{k}}$ (i.e., a probability distribution on agent
types $t$ and amounts of money $i$) with certain properties.  We now
define the appropriate $q$.
Let
\begin{equation}
\label{eqn:q}
q(t,i)~=~(\omega_t)^i~/~\left(\sum_t\sum_{j = 0}^{k_t}~(\omega_t)^j\right).
\end{equation}


It is not immediately clear why this is the right choice of $q$.  As the
following lemma shows, this definition allows us to
characterize the distribution that maximizes the probability of being
in a state corresponding to that distribution (as given by
Lemma~\ref{lem:count})
in terms of relative entropy.

\lem \label{lem:q}
The unique maximum of
$V(d) = H(d)-H(\vec{f}) - \log Z + \sum_t \sum_{i = 0}^{k_t} i d^t_i \log \omega_t$
on $\Delta_{\vec{f},m,\vec{k}}$ occurs at $d^*_q$.
\elem

\prf
For brevity, we drop the superscript $\vec{x}$ on $d$
and let $Y = \sum_t \sum_j (\omega_t)^j$.

\begin{eqnarray*}
{\argmax}_d V(d)
& = & {\argmax}_d (H(d) - H(\vec{f}) - \log Z + \sum_t \sum_{i = 0}^{k_t}
i d(t,i) \log \omega_t)\\
& = & {\argmax}_d (H(d) + \sum_t \sum_{i = 0}^{k_t}
i d(t,i) \log \omega_t)\\
& = & {\argmax}_d \sum_t \sum_{i = 0}^{k_t} [-d(t,i) \log d(t,i)
+ i d(t,i) \log \omega_t]\\
& = & {\argmax}_d \sum_t \sum_{i = 0}^{k_t}[ -d(t,i) \log d(t,i)
+ d(t,i) \log (q(t,i)Y)]\\
& = & {\argmax}_d \sum_t \sum_{i = 0}^{k_t}[ -d(t,i) \log d(t,i)
+ d(t,i) \log q(t,i) + d(t,i)\log Y]\\
& = & {\argmax}_d \log Y + \sum_t \sum_{i = 0}^{k_t}[ -d(t,i) \log d(t,i)
+ d(t,i) \log q(t,i)]\\
&= & {\argmin}_d \sum_t \sum_{i = 0}^{k_t} [d(t,i) \log d(t,i)
- d(t,i) \log q(t,i)]\\
&= & {\argmin}_d \sum_t \sum_{i = 0}^{k_t} d(t,i) \log
\frac{d(t,i)}{q(t,i)}\\
& = & {\argmin}_d H(d || q).
\end{eqnarray*}

By definition, $d^*_q$ minimizes $H(d || q)$.
It is  unique  because $H$ (and thus $V$) is a strictly concave
function on a closed convex set.
\eprf

Lemma~\ref{lem:q} tells us that the most likely distributions of money
to be observed are those with low relative entropy to $q$.
%
Among all distributions in $\Delta_{\vec{f},m,k}$, relative entropy is
minimized by $d^*_q$.  However,
given $n$, it is
quite possible that $d^*_q$ is not $d^{\vec{x}}$ for any $\vec{x}$.
For example, if $d^*_q(t,i) = 1/3$ for
some $t$ and $i$, but $f_thn = 16$, then $d^{\vec{x}}(t,i) = d^*_q(t,i)$
only if
exactly $16/3$ agents of type $t$ to have $i$ dollars,
which cannot be the case.
However, as
the following lemma shows, for sufficiently large $n$, we can always
find a $d^{\vec{x}}$ that is arbitrarily close to $d^*_q$.  For
convenience, we use the 1-norm as our notion of distance.

\lem \label{lem:rounding}
For all $\epsilon$, there exists  $n_\epsilon$ such that, if $n >
n_\epsilon$, then for some state $\vec{x}$, $||d^{\vec{x}} - d^*_q|| <
\epsilon$.
\elem

\prf
Given $n$, we construct $d \in \Delta_{\vec{f},m,k}$ that is of the form
$d^{\vec{x}}$ and is close to $d^*_q$ in a number of steps.  As a first
step, for all $t$ and $i$, let $d_1(t,i)$ be the result of rounding
$d^*_q(t,i)$ to the nearest $1 / hn$ (where ties are broken
arbitrarily).  The function $d_1$ may not be in $\Delta_{\vec{f},m,k}$;
we make minor adjustments to it to get a function in
$\Delta_{\vec{f},m,k}$.  First, note that we may have
$\sum_{i} d_1(t,i) \neq f_t$.  Since $f_t$ is a multiple of $1/hn$,
can get a function $d_2$ that satisfies these constraints by modifying
each term $d_1(t,i)$ by either adding $1/hn$ to it, subtracting $1/hn$
from it, or leaving it alone.
Such a function $d_2$
may still violate the final constraint that $\sum_{t,i} id_2(t,i) = m$.
We construct a function $d_3$ that satisfies this constraint (while
continuing to satisfy the constraint that $\sum_i d_3(t,i) = f_t$)
as follows.  Note that if we increase $d_2(t,i)$ by $1/hn$ and decrease
$d_2(t,j)$ by $1/hn$, then we keep the keep $\sum_i d_2(t,i) = f_t$, and
change $\sum_i i d_2(t,i)$ by $(i-j)/hn$.
Since each term
$d_2(t,i)$ is a multiple of $1 / hn$ and $m$ is a multiple of $1 / h$,
we can perform these adjustments until all the constraints are satisfied.


The rounding to create $d_1$ changed each $d_1(t,i)$ by at most
$1/hn$, so $||d^*_q - d_1||_1 \leq (\sum_t k_t + 1) / hn$.  Since,
each term $d_1(t,i)$ was changed by at most $1/hn$ to obtain $d_2(t,i)$,
we have
$||d_1 - d_2||_1 \leq (\sum_t k_t + 1) / hn$.
Let $c = \max_t(\max(k_t - m, m))$.  Each movement of up to $1 / hn$
in the creation of $d_1$ and $d_2$ altered $m$ by at most $c / hn$.
Thus at most $2c$ movements are needed in the creation of $d_3$ for
each pair $(t,i)$.
Therefore,
$||d_2 - d_3||_1 \leq (\sum_t k_t + 1)2c / hn$.
By the triangle inequality,
$||d^*_q - d_3|| \leq (\sum_t k_t + 1)(2c+2) / hn$, which is
$O(1/n)$.  Hence, for $n_\epsilon$ sufficiently large, the resulting
$d_3$
will always be within distance $\epsilon$ of $d^*_q$.

Finally, we need to show that $d_3 = d^{\vec{x}}$ for some $\vec{x}$.
Each $d_3(t,i)$ is a multiple of $1 / hn$.  There are $hn$ agents in
total, so we can find such an $\vec{x}$ by taking any allocation of
money such that $d_3(t,i)hn$ agents of type $t$ have $i$ dollars.
\eprf

We are now ready to prove Theorem~\ref{thm:distribution}.  We repeat
the statement here for the reader's convenience.

\rethm{thm:distribution}
For all games $(T,\vec{f},h,m,1)$, all vectors $\vec{k}$ of thresholds,
and  all $\varepsilon > 0$,
there exist $q \in \Delta_{\vec{f},m,\vec{k}}$ and $n_\varepsilon$ such that,
for all $n > n_\varepsilon$,
there exists a round $r^*$  such that, for all $r > r^*$,
we have $\Pr(I^r_{q,n,\varepsilon} = 1) > 1-\varepsilon$.
\erethm

\prf From Lemma~\ref{lem:stationary}, we know that, after a
sufficient amount of time, the probability of being in state $\vec{x}$
will be close to $\pi_{\vec{x}} = w_{\vec{x}} / Z$.
Since $\M$ converges to a limit distribution, it is sufficient to show
that the theorem holds in the
limit as $r~\rightarrow~\infty$.
If the theorem holds in the limit for
some $\varepsilon' < \varepsilon$,
then we can take $r$ large enough that the L1 distance
between the
distribution of the chain at time $r$ and the
limit distribution
(i.e. treating the distributions as vectors and computing
the sum of the absolute values of their differences)
is at most $\varepsilon - \varepsilon'$.

The remainder of the proof is essentially that of Theorem 3.13 in
\cite{MaxEnt} (applied in a very different setting).
Let $V(d) = H(d) - H(\vec{f}) - \log Z +
\sum_t \sum_{i = 0}^{k_t} i d(t,i) \log \omega_t$.
We show there exists a value $v_L$ such that, for all states $\vec{x}$
such that $d^{\vec{x}}$ is not within $\varepsilon$ of $d^*_q$,
we have
$V(d^{\vec{x}}) \leq v_L$, and a value $v_H > v_L$ such that
$v_H = V(d^{\vec{y}})$ for some point $\vec{y}$ such that $d^{\vec{y}}$
is within distance $\varepsilon$ of $d^*_q$.
Lemma~\ref{lem:count} then shows that it is
exponentially more likely that $d^{\vec{x}^r} = d^{\vec{y}}$ than
\emph{any} distribution $d$ such that $V(d) \leq v_L$.
If $\vec{x}^r = \vec{y}$ then $I^r_{q,n,\varepsilon} = 1$, and if
$I^r_{q,n,\varepsilon} = 0$ then $V(d^{\vec{x}^r}) \leq v_L$, so this
suffices to establish the theorem.

By Lemma~\ref{lem:q}, the unique maximum of $V$ on
$\Delta_{\vec{f},m,\vec{k}}$ occurs at $d^*_q$.
The set $\{ d \in \Delta_{\vec{f},m,\vec{k}} \mid ||d^*_q - d||_2 \geq
\varepsilon \}$ is closed.
$V$ is a continuous function, so it
takes some maximum $v_L$ on this set.  Pick some $v_H$ such that
$v_L < v_H < V(d^*_q)$.  By the continuity of $V$, there exists an
$\epsilon$ such that if $||d^*_q - d||_1 < \epsilon$ then $V(d)
\geq v_H$.  By Lemma~\ref{lem:rounding}, for sufficiently large $n$,
there is always some $\vec{x}$ such that
$||d^*_q - d^{\vec{x}}||_1 < \epsilon$.
Thus, for some $\vec{x} \in X_{\varepsilon,q}$,
$V(d^{\vec{x}}) \geq v_H$.

$\Pr(I^r_{qn,,\epsilon} = 1) \geq \Pr(\vec{x}^r \in
\{ \vec{y} \mid d^{\vec{y}} = d^{\vec{x}} \})$.
By Lemma~\ref{lem:count},
$\Pr(I^r_{q,n,\epsilon} = 1)$ is at least
$1/F(hn)e^{hnV(d^{\vec{x}})} \geq 1/F(hn)e^{hnv_H}$.
Now consider a $\vec{y}$ such that
$I_{q,n,\epsilon}(\vec{y}) = 0$.  By Lemma~\ref{lem:count}, the
probability that $d^{\vec{x}^r} = d^{\vec{y}}$ is at most
$G(hn)e^{hnV(d^{\vec{y}})} \leq G(hn)e^{hnv_L}$.
There are at most $(hn+1)^{\sum_t (k_t + 1)}$ such points, a number
which is polynomial in $hn$.  Thus, for
$G'(hn) = G(hn)(hn+1)^{\sum_t (k_t + 1)}$,
the probability that $I^r_{q,n,\epsilon} = 0$ is at most
$G'(hn)e^{hnv_L}$.
The ratio of these probabilities is at most
%
$$\frac{G'(hn) e^{hn v_L}}{\frac{1}{F(hn)} e^{hn v_H}}
= \frac{G'(hn)F(hn)}{e^{hn(v_H - v_L)}}.$$
This is the ratio of a polynomial to an exponential, so the
probability of seeing a distribution of distance greater than
$\varepsilon$  from $d^*_q$ goes to zero as $n$ goes to
infinity.
\eprf

\section{Proofs from Section~\ref{sec:strategic}}
\label{sec:threshold}

In this appendix, we provide the omitted proofs from
Section~\ref{sec:strategic}.

The proof of Theorem~\ref{thm:threshold}
relies on modeling the game from the
perspective of a single agent.
Consider a vector $\vec{k}$ of thresholds and the corresponding strategy
profile $\vec{S}(\vec{k})$.  Fix an agent $i$ of type $t$.  Assume that
all the agents other than $i$ continue play their part of
$\vec{S}(\vec{k})$.  What is $i$'s best response?  Since the set of
agents is large, $i$'s choice of strategy will have (essentially) no
impact on the distribution of money.
By Theorem~\ref{thm:distribution}, the distribution of
money will almost always be close to a distribution $d^*$.  Suppose,
the distribution were exactly $d^*$.  Since we know the exact
distribution of money and the thresholds used by the other agents, we
can calculate the number of each type of agent that wish to
volunteer and thus the probabilities that our single agent will be
able to earn or spend a dollar.
Thus, by assuming the distribution
of money is always exactly $d^*$, we can model the game from the
perspective of agent $i$
as a Markov Decision Process (MDP).  We
show in Lemma~\ref{lem:subadditive}
that this MDP has an optimal threshold policy. (Threshold
policies are known as \emph{monotone} policies in the more general
setting where there are more than two
actions.)  We then prove that any optimal policy for the MDP is an
$\epsilon$-best reply to the strategies of the other agents in the
actual game.

\sloppypar{
Taking notation from Puterman~\citeyear{puterman},
we formally define the MDP $\mathcal{P}_{G,\vec{S}(\vec{k}),t}
= \mbox{$(S,A,p(\cdot \mid s,a),r(s,a))$}$ that describes the game where
all the
agents other than $i$ are playing $\vec{S}(\vec{k})_{-i}$ and $i$ has
type $t$.
}
\begin{itemize}
\item $S =
\{0, \ldots , mhn \}$ is the set of possible states for the MDP (i.e., the
possible amounts of money compatible with the distribution $d^*$).
\item $A = \{0 , 1 \}$ is the set of possible actions
for the agent,
where 0 denotes not volunteering and 1 denotes volunteering
iff another agent who has at least one dollar makes a request.
\item $p_u$ is the probability of earning a dollar, assuming the agent
volunteers
(given that all other agents have fixed their thresholds according to
$\vec{k}$ and the distribution of money is exactly
$d^*$).  Each agent of type $t'$ who wishes to volunteer can do so with
probability $\beta_{t'}$.  Assuming exactly the expected number of
agents are able to volunteer,
$\upsilon_{t'} = \beta_{t'}(f_{t'} - d^*(t',k_{t'}))n$
agents of type $t'$ volunteer.  Note that we are disregarding the
effect of $i$ in computing the $\upsilon_{t'}$, since this will have a
negligible effect for large $n$.
Using the $\upsilon_t$s, we can express $p_u$ as the product of two
probabilities: that some agent other than $i$ who has a dollar is
chosen to make a request and that $i$ is the agent chosen to satisfy
it. Thus,
\begin{equation}
\label{eqn:pu}
p_u = \left(\sum_{t'} \rho_{t'}( f_{t'} - d^*(t',0))\right)
\left(\frac{\chi_t\beta_t}{\sum_{t'}\chi_{t'}\upsilon_{t'}}\right).
\end{equation}
\item $p_d$ is the probability of agent $i$ having a request satisfied, given
that agent $i$ has a dollar.
Given that all agents are playing a threshold strategy, if the total
number $n$ of agents is sufficiently large, then it is almost certainly
the case that some agent will always be willing and able
to volunteer.  Thus, we can take $p_d$ to be the probability that agent
$i$ will be chosen to make a request; that is,
\begin{equation}
\label{eqn:pd}
p_d = \frac{\rho_t}{n}
\end{equation}
\item $r(s,a)$ is the (immediate) expected reward for performing
action $a$ in state $s$.  Thus, $r(s,0) = \gamma_t p_d $ if $s > 0$;
$r(0,0) = 0$; $r(s,1) = \gamma_t p_d - \alpha_t p_u$ if $s > 0$; and
$r(0,1) = - \alpha_t p_u$.
\item $p(s' \mid s,a)$ is the probability of being in state $s'$ after
performing action $a$ in state $s$;
$p(s' \mid s,a)$ is determined by $p_u$ and $p_d$; specifically,
$p(s+1 \mid s,1) = p_u$, $p(s-1 \mid s,a) = p_d$ if $s > 0$, and
the remainder of the probability is on $p(s \mid s,a)$
(i.e., $p(s \mid s,a) = 1 - (p(s+1 \mid s,1) + p(s-1 \mid s,a)$).
\item $u^*(s)$ is the expected utility of being in state $s$
if agent $i$ uses the optimal policy for the MDP
$\mathcal{P}_{G,\vec{S}(\vec{k}),t}$
\item $u(s,a)$ is the expected utility for performing action $a$ in
state $s$, given that the optimal strategy is followed after this
action;
$$u(s,a) = r(s,a) + \delta \sum_{s'=0}^{mhn} p(s' \mid s,a) u^*(s').$$
\end{itemize}

To prove Theorem~\ref{thm:threshold}, we need two preliminary lemmas
about the MDP
$\mathcal{P}_{G,\vec{S}(\vec{k}),t}$.

\lem \label{lem:concave}
For the MDP $\mathcal{P}_{G,\vec{S}(\vec{k}),t}$,
$u^*(s+2) + u^*(s) \leq 2 u^*(s + 1)$.
\elem

\prf
The MDP $\mathcal{P}_{G,\vec{S}(\vec{k}),t}$ has an optimal stationary
policy~\cite[Theorem 6.2.10]{puterman} (a policy where the chosen
action depends only on the current state).
Let $\pi$ be such a policy.
Consider the policy $\pi'$
starting in state $s+1$ that ``pretends'' it actually started in state $s$
and is following $\pi$.
More precisely, if $s_0 = s+1$ and $s_j > 0$ for $j = 0, \ldots, k$,
define $\pi'(s_0, s_1, \ldots, s_k) = \pi(s_k - 1)$;
otherwise, if $j \le k$ is the least index such that $s_j = 0$, define
$\pi'(s_0, \ldots, s_k) = \pi(s_k)$.
Given a history $(s_0, \ldots, s_k)$, $j$ is the random variable whose
value is the minimum $i$ such that $s_i = 0$ or $\infty$ if no such
value exists.
The definition of $\pi'$ from $\pi$ creates a bijection between
histories that start in state $s+1$ and histories that start in state $s$,
such that if $h'$ corresponds to $h$,
the probability of history $h'$ with policy $\pi'$ is the same as the
probability of $h$ with policy $\pi$.
Technically, making the mapping a bijection
requires the introduction of a new state $0'$, which
intuitively represents the state where the agent has zero dollars
and missed an opportunity to have a request satisfied last round
because of it.  More formally, we let $p(0' \mid 0,a) = p_d$ and
$p(s \mid 0' , a) = p(s \mid 0,a)$.  With this change, the
probabilities of corresponding histories are the same because
the probability of transitioning from a state to the one
``immediately below'' it (where $s-1$ is immediately below $s$, $0'$ is
immediately below 0, and $0'$
is immediately below itself) is always $p_d$, and the probability of
transitioning
from a state to the one ``immediately above'' it (where $s+1$ is immediately
above $s$, and
$1$ is immediately above $0'$) is always $p_u$.%
\footnote{Note that this means that $0'$ is immediately below $0$ but
$1$ is
immediately above $0'$.  This is intended, because $0'$ intuitively
represents the state where the agent has 0 dollars and had a request
go unsatisfied due to a lack of money in the previous round, so if he
then earns a dollar he will have 1 dollar regardless of whether or not
his request of two rounds previous was satisfied.}

This argument shows that an agent starting with $s+1$ dollars
``pretending'' to start with $s$ will have the same expected reward
each round as an agent who actually started with $s$ dollars, except
during the first round $j$ in a history such that $s_j = 0$.  Thus
(treating $j$ as a random variable), we have
$$u^*(s+1) \geq u^*(s) + E[\delta^j \gamma_t].$$

Similarly, we can use $\pi$ starting from state $s+2$ to define a policy
$\pi''$ starting from state $s+1$, where $i$ ``pretends'' he has one
more dollar and is using $\pi$, up to the first round $j'$ that he is
chosen to make a request with $\pi$ in a state where he has no money (in
which case he can make the request with $\pi$ started from $s+2$, but
cannot make it with $\pi''$ started from $s+1$); from that
point on, he uses $\pi$.
For corresponding histories,
the utilities of an agent starting with $s+1$ dollars and following
$\pi''$ and an agent starting with $s+2$ dollars and following $\pi$
will be the same, except during round $j'$ the agent following $\pi$
will have a request satisfied but the agent following $\pi''$ will not.
Thus,
$$u^*(s+1) \geq u^*(s+2) - E[\delta^{j'} \gamma_t].$$

Since, if $i$ uses $\pi$, he will run out of money sooner if he starts
with $s$ dollars than if he starts with $s+2$ dollars,
$$E[\delta^j \gamma_t] > E[\delta^{j'} \gamma_t].$$
Thus, $u^*(s+2) + u^*(s) \leq 2 u^*(s + 1)$.
\eprf

\lem \label{lem:subadditive}
$\mathcal{P}_{G,\vec{S}(\vec{k}),t}$ has an optimal threshold policy.
\elem

\prf
As shown by Puterman~\citeyear[Lemma 4.7.1]{puterman}, it suffices to
prove that $u(s,a)$ is subadditive.
That is, we need to prove that, for all states $s$,
\begin{equation}\label{eq1}
u(s+1,1) + u(s,0) \leq u(s+1,0) + u(s,1).
\end{equation}
We consider here only the case that $s > 0$ (the argument is
essentially the same if $s = 0$).
Because $s > 0$, $r(s+1,a) = r(s,a)$, so
(\ref{eq1}) is equivalent to
$$
\begin{array}{ll}
&p_u u^*(s+2) + p_d u^*(s) + (1 - p_u - p_d) u^*(s+1)
+ p_d u^*(s-1) + (1 - p_d) u^*(s)\\
\leq &p_d u^*(s) + (1 - p_d) u^*(s+1)
+ p_u u^*(s+1) + p_d u^*(s-1) + (1 - p_u - p_d) u^*(s).
\end{array}
$$
This simplifies to
$$u^*(s+2) + u^*(s) \leq 2 u^*(s + 1),$$
which follows from Lemma~\ref{lem:concave}.
\eprf

We can now prove Lemma~\ref{lem:br} and Theorem~\ref{thm:threshold}.

\relem{lem:br}
Consider the games $G_n = (T,\vec{f},h,m,n)$
(where $T$, $\vec{f}$, $h$, and $m$ are fixed, but $n$ may vary).
There exists a $k$ such that for all $n$, $s_k$ is
an optimal policy for $\mathcal{P}_{G_n,\vec{S}(\vec{k}),t}$.
The threshold
$k$ is the maximum value of $\kappa$ such that
\begin{equation}
\tag{\ref{eqn:policy}}
\alpha_t \leq E[(1 - (1 - \delta_t)/n)^{J(\kappa,p_u,p_d)}] \gamma_t,
\end{equation}
where $J(\kappa,p_u,p_d)$ is a random variable whose value is the first
round in which an agent starting with $\kappa$ dollars,
using strategy $s_{\kappa}$, and with probabilities $p_u$ and $p_d$ of earning
a dollar and of being chosen given that he volunteers,  respectively,
runs out of money.
\erelem

\prf
Fix $n$.
Suppose that an agent is choosing between a threshold of $\kappa$ and a
threshold of $\kappa+1$.  These policies only differ when the agent
has $\kappa$
dollars: he will volunteer with the latter but not with the former.
If he volunteers when he has $\kappa$ dollars and is chosen, he will
pay a cost of $\alpha_t$ and he will have $\kappa+1$ dollars.  As in
the proof of Lemma~\ref{lem:concave}, we can define a bijection on
histories such that, in corresponding histories of equal probability,
an agent who started with $\kappa$ dollars and is using $s_{\kappa}$
will always
have one less dollar than an agent who started with $\kappa+1$ dollars and
is using $s_{\kappa+1}$, until the first round $r$ in which the agent using
$s_{\kappa+1}$ has zero dollars.  This means that in round $r-1$ the
agent using $s_{\kappa+1}$ had a request satisfied but the agent using
$s_k$ was unable
to because he had no money.  Thus, if the agent volunteers when he has
$\kappa$ dollars and pays a cost of $\alpha_t$ in the current round, the
expected value of being able to spend that dollar in the future is
$E[(1 - (1 - \delta_t)/n)^{J(\kappa+1,p_u,p_d)}] \gamma_t$.  Since this
expectation is
strictly increasing in $\kappa$ (an agent with more money takes longer to
spend it all), the maximum $\kappa$ such that Equation~(\ref{eqn:policy})
holds is an optimal threshold policy.

Taking the maximum value of $\kappa$ that satisfies
Equation~(\ref{eqn:policy}) ensures that, for the $n$ we fixed, we
chose the maximum optimal threshold.  We now need to show that this
maximum optimal threshold is independent of $n$, which we do by
showing that the expecting utility of every threshold policy $s_k$ is
independent of $n$.  The expected utility of a policy depends on the
initial amount of money, but since an agent's current amount of money
is a random walk whose transition probabilities are determined by
$p_u$ and $p_d$, there is a well-defined limit probability
$$x_i^* =  \lim_{r \rightarrow \infty} \Pr(
\mbox{agent has } i \mbox{ dollars in round } r)$$
determined by the ratio $p_u / p_d$
(this is because the limit distribution satisfies the detailed
balance condition: $x_i^* p_u = x_{i+1}^* p_d$).
This distribution has the property that if the
agent starts with $i$ dollars with probability $x_i^*$,
then in every round the probability he has $i$ dollars is
$x_i^*$.  Thus, in each round his expected utility is
$\gamma p_d (1 - x_0^*) - \alpha p_u (1 - x_k^*)$.
We can factor out $n$ to write $p_u = p_u'/n$ and $p_d = p_d'/n$ where
$p_u'$ and $p_d'$ are independent of $n$.  Note that $p_u / p_d = p_u' /
p_d'$, so the $x_i^*$'s are independent of $n$.  Thus, we can rewrite the
agent's expected utility for each round as $c / n$, where
$c = \gamma p_d' (1 - x_0^*) - \alpha p_u' (1 - x_k^*)$ is
independent of $n$.
Therefore, the expected
utility of $s_k$ is
$$\sum_{r = 0}^{\infty} \left(1 - \frac{1 - \delta_t}{n}\right)^r
\frac{c}{n} = \frac{c}{1 - \delta_t},$$
which is independent of $n$.
\eprf

\rethm{thm:threshold}
For all games $G = (T,\vec{f},h,m,n)$, all vectors $\vec{k}$ of
thresholds, and  all $\varepsilon > 0$,
there exist $n^*_\varepsilon$ and $\delta^*_{\varepsilon,n}$
such that for all $n > n^*_\varepsilon$,
types $t \in T$, and $\delta_t > \delta^*_{\varepsilon,n}$,
an optimal threshold policy for
$\mathcal{P}_{G,\vec{S}(\vec{k}),t}$ is an
$\varepsilon$-best reply
to the strategy profile $\vec{S}(\vec{k})_{-i}$
for every agent $i$ of type $t$.
\erethm

\prf
By Lemma~\ref{lem:subadditive},
$\mathcal{P}_{G,\vec{S}(\vec{k}),t}$ has an optimal threshold policy.
However, this might not be
a best reply  for agent $i$ in the actual game if the other agents
are playing $\vec{S}(\vec{k})$.
$\mathcal{P}_{G,\vec{S}(\vec{k}),t}$
assumes that the probabilities
of earning or spending a dollar in a given round
are always exactly $p_u$ and $p_d$ respectively.
Theorem~\ref{thm:distribution} guarantees only that, in the game, the
corresponding probabilities are close to $p_u$ and $p_d$ with high
probability after some amount of time that can depend on $n$.
A strategy $S$ for player $i$ in $G$ defines a policy $\pi_S$ for
$\mathcal{P}_{G,\vec{S}(\vec{k}),t}$
in the obvious way; similarly, a policy for the MDP determines a
strategy for player $i$ in the game.
The expected utility of $\pi_S$ is close to
$U_i(S,\vec{S}(\vec{k})_{-i})$, but is, in general,
not equal to it, because, as we noted, $p_u$ and $p_d$ may differ from
the corresponding probabilities in the game.  They differ for three reasons:
(1) they are close, but not identical; (2) they are only close with high
probability, and (3) they are only close after some amount of time.
As we now show, given $\varepsilon$, the difference in the expected
utility due to each reason can be bounded by $\varepsilon / 6$,  so
the expected utility of any strategy is within
$\varepsilon / 2$ of the value
the corresponding policy in $\mathcal{P}_{G,\vec{S}(\vec{k}),t}$.
Thus, an optimal strategy for the MDP is an $\varepsilon$-best reply.

As we have seen, the probabilities $p_u$ and $p_d$ are determined by the
number of agents of each type that volunteer
(i.e., the expressions $\upsilon_{t'}$ for each type $t'$).
The distance between $d^{\vec{x}^r}$ and
$d^*$ bounds how much the actual number of agents of type $t'$ that
wish to volunteer in round $r$ can differ from
$\upsilon_{t'} / \beta_{t'}$.
Even if exactly $\upsilon_{t'} / \beta_{t'}$ agents wish to volunteer
for each type $t'$, there might not be exactly $\upsilon_{t'}$ agents
who actually volunteer because of the stochastic decision by nature
about who can volunteer and because $i$ cannot satisfy his own
requests.
However, for sufficiently large $n$, the effect on $p_u$ and $p_d$
from these two factors is arbitrarily close to zero.
Applying Theorem~~\ref{thm:distribution}, there exist $n_1$ and $r_1$
such that if there are at least $n_1$ agents, for all round $r > r_1$,
$d^{\vec{x}^r}$ and $d^*$ are sufficiently close that
the difference between the utility of policy $\pi_{S'}$ in the MDP and
$U_i((S',\vec{S}_{-i})$
in rounds $r > r_1$ where $d^*$ is sufficiently close is at most
$\varepsilon / 6$.

Note that
the maximum possible difference in utility between a round of the
MDP and a round of the game is
$\gamma + \alpha$ (if agent $i$ spends a dollar rather than earning
one).
Applying Theorem~\ref{thm:distribution} again,
for $e = \varepsilon/6(\gamma + \alpha)$,
there exist $n_2$ and
$r_2$ such that the probability of the distribution not being within
$e$ of $d^*$ is less than $e$.
Thus, the difference
between the expected utility of policy $\pi_{S'}$ in the MDP and
$U_i((S',\vec{S}_{-i})$
in rounds $r > r_2$ where $d^*$ is not sufficiently close is at most
$e(\gamma + \alpha) = \varepsilon/6$.

Let
$n_\varepsilon^* = \max(n_1,n_2)$ and
$r^* = \max(r_1,r_2)$.
The values of $n^*_\varepsilon$ and $r^*$ do not depend on $\delta$,
so we can take $\delta^*_{\varepsilon,n}$ to be sufficiently close to 1
that the total utility from the first $r^*$ rounds is at most
$\varepsilon / 6$, completing the proof of the theorem.
\eprf

Recall that $\BR_G$ maps a vector $\vec{k}$ describing the threshold
strategy for each type to a vector $\vec{k}'$ of best replies.

\relem{lem:monotone}
Consider the family of games $G_m = (T,\vec{f},h,m,n)$ and the
strategies $\vec{S}(\vec{k})$, for $mhn < \sum_t f_t k_t hn$.
For this family of game,
$\lambda_{m,\vec{k}}$ is non-decreasing in $m$ and non-increasing in
$\vec{k}$; $p_u^{m,\vec{k}}$ is non-decreasing in
$m$ and non-increasing in $\vec{k}$; and the function
$\BR_G$ is non-decreasing in $\vec{k}$ and non-increasing in
$m$.
\erelem

\prf
We first show that that $\lambda_{m,\vec{k}}$ is monotone in $m$ and
$\vec{k}$.  We then show that $p_u^{m,\vec{k}}$ is a monotone function
of $\lambda_{m,\vec{k}}$ and that $\BR_G$ is a monotone
function of $p_u^{m,\vec{k}}$, completing the proof.

We now show that $\lambda_{m,\vec{k}}$ is non-decreasing in $m$.
Fix a vector of thresholds $\vec{k}$ and let
\begin{equation}
\label{eqn:g}
g_{\vec{k}}(\lambda) = \sum_{t,i} i \frac{f_t \lambda^i q_{\vec{k}}(t,i)}
{\sum_{j = 0}^{k_t} \lambda^j q_{\vec{k}}(t,j)},
\end{equation}
where $q_{\vec{k}}$ is the value of $q$ from Equation~(\ref{eqn:q})
(we add the subscript $\vec{k}$ to stress the dependence on $\vec{k}$).
The definition of $\lambda_{m,\vec{k}}$ in Equations (\ref{eqn:d}) and
(\ref{eqn:m}) in
Lemma~\ref{lem:minrelent} ensures that, for all $m$, $m =
g_{\vec{k}}(\lambda_{m,\vec{k}})$.
A relatively straightforward computation shows that
$g'_{\vec{k}}(\lambda) > 0$ for all $\lambda$.
Thus, if $m' > m$, $g_{\vec{k}}(\lambda) = m$, and
$g_{\vec{k}}(\lambda') = m'$, we must have $\lambda' > \lambda$.
It follows that $\lambda_{m,\vec{k}}$ is increasing in $m$.
(Note that $\lambda_{m,\vec{k}}$ is undefined for
$m \geq \sum_t f_t k_t$, which
is why monotonicity holds only
for
values of $m$ such that
$mhn < \sum_t f_t k_t$.)

We next show that $\lambda_{m,\vec{k}}$ is non-increasing in
$\vec{k}$.
Since we have a finite set of types, it suffices to consider the case
where a single type $t^*$ increases its threshold by 1.
Let $\vec{k}$ denote
the initial vector of thresholds, and let $\vec{k}'$ denote the vector
of thresholds after agents of type $t^*$ increase their threshold by
1; that is, $k_t = k'_t$ for $t \ne t^*$, and $k'_{t^*} = k_{t^*} + 1$.

The first step in showing that $\lambda_{m,\vec{k}}$ is non-increasing in
$\vec{k}$ is to show that $g_{\vec{k}'}(\lambda_{m,\vec{k}}) >
g_{\vec{k}}(\lambda_{m,\vec{k}}) = m$.
We do this by breaking the sum in the definition of $g$ in
Equation~(\ref{eqn:g}) into two pieces; those terms where $t \ne t^*$,
and those where $t=t^*$.

It follows immediately from Equation (\ref{eqn:q}) that
there exists a constant $c$ such that, for all $i$ and $t \ne t^*$, we have
$q_{\vec{k'}}(t,i) = c q_{\vec{k}}(t,i)$.
It follows from Equation~(\ref{eqn:d}) that
for all $i$ and $t \ne t^*$, since $k_t = k'_t$, we have
\begin{equation}
\label{eqn:g1}
i \frac{f_t \lambda_{m,\vec{k}}^i q_{\vec{k}'}(t,i)}
{\sum_{j = 0}^{k_t'} \lambda_{m,\vec{k}}^j q_{\vec{k}'}(t,j)}
= i \frac{f_t \lambda_{m,\vec{k}}^i c q_{\vec{k}}(t,i)}
{\sum_{j = 0}^{k_t'} \lambda_{m,\vec{k}}^j c q_{\vec{k}}(t,j)}
= i \frac{f_t \lambda_{m,\vec{k}}^i  q_{\vec{k}}(t,i)}
{\sum_{j = 0}^{k_t} \lambda_{m,\vec{k}}^j q_{\vec{k}}(t,j)};
\end{equation}
that is, the corresponding terms in the sum for
$g_{\vec{k}'}(\lambda_{m,\vec{k}})$ and
$g_{\vec{k}}(\lambda_{m,\vec{k}})$ are the same if $t \ne t^*$.

Now consider the corresponding terms for type $t^*$.  First observe that
for all $i < k_t'$,
\begin{equation}\label{neweq1}
\frac{f_{t^*} \lambda_{m,\vec{k}}^i q_{\vec{k}'}(t^*,i)}
{\sum_{j = 0}^{k_{t^*}'} \lambda_{m,\vec{k}}^j q_{\vec{k}'}(t^*,j)}
< \frac{f_{t^*} \lambda_{m,\vec{k}}^i  q_{\vec{k}}(t^*,i)}
{\sum_{j = 0}^{k_{t^*}} \lambda_{m,\vec{k}}^j q_{\vec{k}}(t^*,j)};
\end{equation}
the two terms have essentially the same numerator
(the use of $q_{\vec{k}'}$ instead of $q_{\vec{k}}$ cancels out as in
Equation~(\ref{eqn:g1})),
but the first has
a larger denominator because $k'_{t^*} = k_{t^*} +1$, so there is one
more term in the sum.
Since $f_{t^*} = \sum_{i = 0}^{k_{t^*}} d^*_{q_{\vec{k}}}(t^*,i) =
\sum_{i = 0}^{k'_{t^*}} d^*_{q_{\vec{k}'}}(t^*,i)$,
by Equations~(\ref{eqn:d}) and~(\ref{eqn:m}),
\begin{equation}\label{neweq2}
\sum_{i = 0}^{k_{t^*}}
\frac{f_{t^*} \lambda_{m,\vec{k}}^i q_{\vec{k}}(t^*,i)}
{\sum_{j = 0}^{k_{t^*}} \lambda_{m,\vec{k}}^j q_{\vec{k}}(t^*,j)}
= \sum_{i = 0}^{k_{t^*}'}
\frac{f_{t^*} \lambda_{m,\vec{k}}^i q_{\vec{k}'}(t^*,i)}
{\sum_{j = 0}^{k_{t^*}'} \lambda_{m,\vec{k}}^j q_{\vec{k}'}(t^*,j)}.
\end{equation}
It follows that
\begin{equation}
\label{eqn:g2}
\sum_{i=0}^{k_{t^*}'}
i \frac{f_{t^*} \lambda_{m,\vec{k}}^i q_{\vec{k}'}(t,i)}
{\sum_{j = 0}^{k_{t^*}'} \lambda_{m,\vec{k}}^j q_{\vec{k}'}(t,j)}
> \sum_{i=0}^{k_{t^*}}
i \frac{f_t \lambda_{m,\vec{k}}^i  q_{\vec{k}}(t,i)}
{\sum_{j = 0}^{k_{t^*}} \lambda_{m,\vec{k}}^j  q_{\vec{k}}(t,j)}.
\end{equation}
To see this, note that
the two expressions above have the form $\sum_{i = 0}^{k_{t^*}+1} i c_i$
and  $\sum_{i=0}^{k_{t^*}} i d_i$, respectively.
By Equation~(\ref{neweq2}), $\sum_{i=0}^{k_{t^*}+1}c_i =
\sum_{i=0}^{k_{t^*}}d_i = f_{t^*}$; by Equation~(\ref{neweq1}),
$c_i < d_i$ for $i = 0,\ldots,k_{t^*}$.  Thus,
in going from the right side to the left side, weight is being
transferred from lower terms to $k_{t^*}+1$.

Combining
Equations~(\ref{eqn:g1}) and~(\ref{eqn:g2}) gives us
$g_{\vec{k}'}(\lambda_{m,\vec{k}}) > g_{\vec{k}}(\lambda_{m,\vec{k}}) = m$,
as desired.
Since $g_{\vec{k}'}(\lambda_{m,\vec{k}'}) = m$, by definition, it
follows that $g_{\vec{k}'}(\lambda_{m,\vec{k}}) >
g_{\vec{k'}}(\lambda_{m,\vec{k}'})$.
Since, as shown above, $g_{\vec{k''}}$ is an increasing function, it
follows that $\lambda_{m,\vec{k}} > \lambda_{m,\vec{k}'}$.
Thus, $\lambda_{m,\vec{k}}$ is decreasing in $\vec{k}$.

We now show that the monotonicity of $\lambda_{m,\vec{k}}$ implies the
monotonicity of $p_u^{m,\vec{k}}$.
To do this, we show that, for all types $t$,
$p_u^{m,\vec{k}} = p_d \lambda_{m,\vec{k}} \omega_t$.
Since $\omega_t$ and $p_d$ are independent of $m$ and $\vec{k}$,
it then follows that the
monotonicity of $\lambda_{m,\vec{k}}$ implies the
monotonicity of $p_u^{m,\vec{k}}$.
(Recall that $\omega_t = \beta_t \chi_t / \rho_t$ was defined in
Section~\ref{sec:model}.)

Fix a type $t'$.  Then, dropping superscripts and subscripts on $p_u$,
$d$, and $\lambda$ for brevity, we have the following sequence of
equalities (where the explanation for some of these lines is given
following the equations):
\begin{align}
\label{eqn:line1}
p_u & =  \left( \sum_t \rho_T (f_t - d(t,0) \right)
\left( \frac{\chi_{t'}\beta_{t'}}{n \sum_t \chi_t \beta_t
(f_t - d(t,k_t))} \right)\\
\label{eqn:line2}
& =  \left(\frac{\sum_t \sum_{i = 1}^{k_t} \rho_t d(t,i)}
{\sum_t \sum_{i = 0}^{k_t-1} \chi_t \beta_t d(t,i)}\right)
\left( \frac{\chi_{t'} \beta_{t'}}{n} \right)\\
\label{eqn:line3}
& =  \left(\frac{\sum_t \sum_{i = 0}^{k_t-1}
\rho_t \lambda \omega_t d(t,i)}
{\sum_t \sum_{i = 0}^{k_t-1} \chi_t \beta_t d(t,i)}\right)
\left( \frac{\chi_{t'} \beta_{t'}}{n} \right)\\
\label{eqn:line4}
& =  \lambda \left(\frac{\sum_t \sum_{i = 0}^{k_t-1}
\chi_t \beta_t d(t,i)}
{\sum_t \sum_{i = 0}^{k_t-1} \chi_t \beta_t d(t,i)}\right)
\left( \omega_{t'} p_d \right)\\
\notag
& =  \lambda \omega_{t'} p_d
\end{align}
Equation~(\ref{eqn:line1}) is just the definition of $p_u$ from
Equation~(\ref{eqn:pu}).
Equation~(\ref{eqn:line2}) follows from the observation that, by
Equation~(\ref{eqn:d}), $f_t = \sum_i d(t,i)$.
Equation~(\ref{eqn:line3}) follows from the observation that, again by
Equation~(\ref{eqn:d}), $d(t,i) = \omega_t \lambda d(t,i-1)$.
Equation~(\ref{eqn:line4}) follows from the definitions of $\omega_t$
and $p_d$ (see Equation~(\ref{eqn:pd})).
Thus, as required,
$p_u^{m,\vec{k}} = p_d \lambda_{m,\vec{k}} \omega_t$.

Finally, we show that the monotonicity of $p_u^{m,\vec{k}}$ implies
the monotonicity of $\BR_G$.
Let $\vec{k}'' = \BR_G(\vec{k})$.
By Lemma~\ref{lem:br},
$k_t''$ is the maximum value of $\kappa$ such that
$$\alpha_t \leq E[(1 - (1 - \delta_t)/n)^{J(\kappa,p_u^{m,\vec{k}},p_d)}]
\gamma_t.$$
We (implicitly) defined the random variable
$J(\kappa,p_u,p_d)$ as a function on histories.
Instead, we can define $J(\kappa,p_u,p_d)$ as a function
on random bitstrings (which intuitively determine a history).  With
this redefinition, it is clear that, if $p_u < p_u'$, for all bitstrings
$b$, we have $J(\kappa,p_u,p_d)(b) < J(\kappa,p_u',p_d)(b)$.  It easily
follows that
$$E[(1 - (1 - \delta_t)^{J(\kappa,p_u',p_d)}]
< E[(1 - (1 - \delta_t)^{J(\kappa,p_u,p_d)}]$$
for all $\kappa$. Thus, the monotonicity of $\BR_G$ follows from the
monotonicity of $p_u^{m,\vec{k}}$.
\eprf

\relem{lem:nontrivial}
For all games $G = (T,\vec{f},h,m,n)$, there exists
a $\delta^*<1$ such
that if $\delta_t > \delta^*$ for all $t$, there is a vector
$\vec{k}$ of thresholds such that $\BR_G(\vec{k}) > \vec{k}$.
\erelem

\prf
Take $\vec{k}$ to be such that $k_t = \lceil m \rceil + 1$ for each
type $t$.  Then by Theorem~\ref{thm:threshold},
there exists a $\vec{k}'$ such that
$\BR_G(\vec{k}) = \vec{k}'$.  By Lemma~\ref{lem:br}, $k_t'$ is
the maximum value of $\kappa$ such that
\begin{equation}
\tag{\ref{eqn:policy}}
\alpha_t \leq
E[(1 - (1 - \delta_t)/n)^{J(\kappa,p_u^{\vec{k}},p_d)}] \gamma_t.
\end{equation}
As $\delta_t$ approaches 1,
$E[(1 - (1 - \delta_t)/n)^{J(\kappa,p_u^{\vec{k}},p_d)}]$
approaches 1, and so the right hand side of
Equation~(\ref{eqn:policy}) approaches $\gamma_t$.
For any standard agent, $\alpha_t < \gamma_t$.
Thus, there exists a $\delta_t$ such that
$$\alpha_t \leq
E[(1 - (1 - \delta_t)/n)^{J(k_t,p_u^{\vec{k}},p_d)}] \gamma_t.$$
For this choice of $\delta_t$, we must have $k_t' \geq k_t+1 > k_t$.
Take $\delta^* = \max_t \delta_t$.
\eprf

\begin{acks}
We would like to thank
Randy Farmer,
Peter Harremoes,
Shane Henderson,
Jon Kleinberg,
David Parkes,
Dan Reeves,
Emin G\"{u}n Sirer,
Michael Wellman,
and anonymous referees and editors
for helpful suggestions, discussions, and criticisms.
Eric
Friedman, Ian Kash, and Joseph Halpern are supported in part by NSF
under grant
ITR-0325453.
Joseph Halpern is
also supported in part by
NSF under grants CTC-0208535, IIS-0812045, and IIS-0534064;
by ONR under grant N00014-01-10-511;
by the DoD Multidisciplinary University Research Initiative (MURI)
program administered by the ONR under grants N00014-01-1-0795 and
N00014-04-1-0725;
and by AFOSR under grants F49620-02-1-0101, FA9550-08-1-0438,
FA9550-05-1-0055,
and FA9550-09-1-0266.
Eric Friedman is also supported in part by NSF under grant CDI-0835706.
\end{acks}

\bibliographystyle{ACM-Reference-Format-Journals}
\bibliography{Z:/Research/Bibliography/kash}

\end{document}